\documentclass[lettersize,journal]{IEEEtran}
\usepackage{cite}
\usepackage{amsmath,amssymb,amsfonts}
\usepackage{algorithmic}
\usepackage{graphicx, color, subfigure}
\usepackage{textcomp}
\usepackage{booktabs} 
\usepackage{array} 
\usepackage{paralist} 
\usepackage{enumitem}
\usepackage{verbatim} 
\usepackage[linesnumbered,ruled]{algorithm2e} 
\usepackage{tabularx}  
\usepackage{engord}
\usepackage{booktabs,multirow}
\usepackage{url}
\usepackage{comment}
\usepackage{nomencl}

\usepackage{float}
\usepackage{mathtools}
\usepackage{bbding}
\usepackage{amsthm}
\usepackage[table]{xcolor} 
\usepackage[T1]{fontenc}

\newcommand{\tabincell}[2]{\begin{tabular}{@{}#1@{}}#2\end{tabular}}
\newcommand\cparagraph[1]{\vspace{0.6mm}\noindent\textbf{#1.}}

\usepackage[affil-it]{authblk}

\begin{document}

\title{Optimization without Future Compromises? Decentralized Coordination via Collective and Reinforcement Learning}

\author{\IEEEauthorblockN{Chuhao Qin\IEEEauthorrefmark{1}, and 
Evangelos Pournaras\IEEEauthorrefmark{1}} \\

\IEEEauthorblockA{
\IEEEauthorrefmark{1}School of Computer Science, University of Leeds, UK
}
        
\thanks{
This research is supported by a UKRI Future Leaders Fellowship (MR-/W009560/1): \emph{Digitally Assisted Collective Governance of Smart City Commons–ARTIO}
Corresponding author: E. Pournaras (email: e.pournaras@leeds.ac.uk).
}
}


\maketitle

\begin{abstract}
Efficient resource allocation in multi-agent systems requires autonomous agents to coordinate their decisions while balancing system-wide objectives with individual costs. This becomes increasingly challenging over long time horizons, where decisions that improve the current allocation may compromise future resource allocation, while decentralized agents have limited observations of the overall system. Multi-agent reinforcement learning (MARL) can learn such long-term dependencies via local observations, but directly applying it to large-scale coordination leads to rapidly growing decision spaces and inefficient training. To this end, we propose Hierarchical Reinforcement and Collective Learning (HRCL), a hierarchical framework that uses MARL to guide, rather than replace, decentralized multi-agent coordination. At the high level, MARL learns strategies that restrict the alternatives considered during coordination and guide agents in balancing system-wide and individual objectives. At the low level, agents perform efficient decentralized coordination under this strategic guidance. This separation reduces the learning space and allows short-term coordination trade-offs to be evaluated according to their long-term effects. Experiments on a synthetic benchmark show that HRCL converges substantially faster than standalone MARL and reduces system-wide and individual costs by 35.53\% and 27.05\%, respectively. Evaluations on energy self-management and drone swarm sensing further show improved resource allocation, power-peak regulation, and sensing efficiency. These results show that learning strategic guidance for an existing coordination process can retain scalable decentralized coordination without letting short-term decisions compromise future resource allocation.
\end{abstract}



\section{Introduction}
Efficient resource allocation is important for building sustainable systems, where limited resources need to be allocated among multiple participants while adapting to changing demands and environments. Multi-agent coordination provides a way for autonomous agents to jointly make such allocation decisions, saving system resources while reducing the costs incurred by individual agents. However, these two objectives are often conflicting~\cite{teghem2000interactive,liefooghe2014hybrid}. For example, households in a smart grid can coordinate their electricity consumption towards periods with lower electricity prices while avoiding system-wide power peaks. Similarly, multiple robots, such as drones, can coordinate their assigned sensing areas to improve the overall efficiency of data collection while reducing the energy consumed by individual robots. Effective coordination therefore requires agents to balance system-wide resource allocation with their individual costs.

Such coordination becomes more challenging when resource allocation extends over a long time horizon. Instead of scheduling electricity consumption within a single short period, households need to coordinate their consumption across multiple days or weeks. Robots also perform multiple tasks over time, where task allocation and recharging decisions in one period influence the remaining tasks in later periods. Directly solving all these decisions together can result in a large combinatorial decision space, especially when the number of agents and available alternatives increases~\cite{barrett2020exploratory}. One approach is to decentralize the coordination, allowing agents to make decisions locally and reducing the computation and information exchange required by a central coordinator~\cite{hinrichs2013cohda}. However, decentralized agents have only limited observations of the overall system. Another approach is to divide a long-term allocation problem into multiple rounds of coordination~\cite{li2021lifelong}. This reduces the decision space at each round, but current decisions then need to account for their effects on future resource allocation. The key challenge is therefore how to efficiently coordinate decentralized agents without letting short-term allocation decisions compromise long-term system objectives.

Multi-agent reinforcement learning (MARL) provides a natural approach to address this challenge by modeling coordination as sequential decision-making~\cite{andrychowicz2016learning,cappart2021combining}. This allows an agent to accept a less favorable decision in the current round when it leads to a better cumulative outcome. Deep neural networks further allow agents to learn coordination policies under complex and changing system conditions. However, directly applying MARL to large-scale decentralized resource allocation introduces another challenge: \textit{how can reinforcement learning efficiently guide decentralized coordination?} As the system scales, more agents and more available alternatives create an increasingly large joint decision space. Directly learning from this space can substantially increase training complexity and slow down convergence~\cite{wong2023deep}. At the same time, decentralized agents make decisions based on limited observations, making it difficult to coordinate their local choices towards system-wide objectives~\cite{wong2023deep}. These problems become more significant for long-horizon allocation, where agents need to consider both a large decision space and the future effects of their current choices.

To address this challenge, this paper proposes Hierarchical Reinforcement and Collective Learning (HRCL), a new hierarchical framework for learning to guide decentralized multi-agent coordination. The key idea is to separate strategic learning from the detailed coordination process. At the low level, an existing decentralized coordination method allows agents to iteratively coordinate their resource allocation decisions with low computational and communication overhead. At the high level, MARL learns coordination strategies from the outcomes of this decentralized process and provides strategic guidance for subsequent coordination. Two complementary strategies are introduced to high-level MARL. The first restricts the set of alternatives considered during coordination, reducing the action space that needs to be learned. The second guides how agents balance system-wide resource allocation and their individual costs. In this way, MARL does not directly solve the entire combinatorial coordination problem. Instead, it learns how to guide a scalable decentralized coordination process over time. This hierarchical design allows low-level coordination to make necessary short-term trade-offs, while high-level learning evaluates these decisions according to their long-term effects.

The contributions of this paper are summarized as follows: (1) A new reinforcement learning framework for guiding decentralized multi-agent coordination. HRCL separates long-term strategic learning from short-term decentralized coordination, allowing MARL to guide repeated coordination decisions according to their cumulative system effects. (2) Two high-level coordination strategies that improve learning efficiency and coordination quality. The plan grouping strategy reduces the high-level action space by restricting the alternatives considered during low-level coordination, while the behavior strategy guides agents in balancing system-wide efficiency and individual discomfort. (3) A hierarchical integration of MAPPO~\cite{yu2022surprising} with the EPOS collective learning approach~\cite{majumdar2023discrete}. At the low level, EPOS coordinates agents through a tree structure and efficiently aggregates their selections into collective outcomes. These outcomes provide system-level feedback for learning high-level coordination strategies, while decentralized execution is retained at the coordination layer. (4) An extensive empirical evaluation of scalability and cross-domain applicability. HRCL is evaluated using a synthetic benchmark and two application-oriented case studies in energy self-management and drone swarm sensing. The experiments show that HRCL outperforms standalone reinforcement learning and decentralized coordination approaches.

\section{Related Work}\label{sec:related_work}
This paper studies how reinforcement learning can provide long-term guidance for scalable decentralized multi-agent coordination. We review related work from three aspects: 

\cparagraph{Decentralized Multi-Agent Coordination for Resource Allocation}
Resource allocation in multi-agent systems requires agents to coordinate coupled decisions while considering both system-wide objectives and their individual costs~\cite{deakin2013smart}. Decentralized coordination is attractive in such settings because agents can make decisions locally and exchange information without relying on a central optimizer. Classical distributed combinatorial optimization methods, such as COHDA~\cite{hinrichs2013cohda} and DPOP~\cite{chen2018bd}, coordinate agents through message passing over a communication structure. Recent studies have also applied distributed optimization to cooperative multi-robot control~\cite{dong2023decentralised} and consensus-based allocation to multi-agent composite tasks~\cite{picard2023multi}. These methods show that decentralized coordination can efficiently distribute resource and task allocation among multiple agents. Furthermore, the collective learning provides an efficient way to aggregate agents' local decisions and adapt their selections through repeated interactions~\cite{pournaras2018decentralized}. EPOS and its multi-objective extension organize agents in a tree structure, where information is aggregated through local message passing instead of requiring every agent to exchange its complete information with all others~\cite{majumdar2023discrete}. This structure allows the coordination process to scale to large agent populations with relatively low computational overhead, making it suitable as the low-level coordination mechanism considered in this work. However, these decentralized approaches mainly focus on determining coordinated decisions within each optimization round. Our work further considers how such coordination can be strategically guided over multiple rounds according to long-term outcomes.

\cparagraph{MARL for Scalable Decentralized Coordination}
Multi-agent reinforcement learning (MARL) provides a natural way to model repeated coordination as sequential decision-making~\cite{tilak2010decentralized,li2020deep,wong2023deep}. By learning from cumulative rewards, agents can account for the long-term effect of their current actions. Centralized training with decentralized execution is widely used in cooperative MARL to exploit global information during training while retaining local policies during execution. Representative approaches include MADDPG~\cite{lowe2017multi} and MAPPO~\cite{yu2022surprising}, where the latter demonstrates that PPO-based learning can provide strong performance across cooperative multi-agent tasks. Rather, a major challenge is scalability. As the number of agents and available decisions increases, the joint decision space grows rapidly, while individual agents have limited observations during decentralized execution~\cite{wong2023deep}. Recent work addresses this problem by reducing communication or improving how global information is used during training~\cite{wu2024collaborative,li2024q,zhang2025neuro}. Sparse communication allows agents to selectively exchange information~\cite{han2023model}, while PTDE distills personalized global information into representations that can be used by local policies during execution~\cite{chen2022ptde}. Other work improves large-scale decentralized reinforcement learning by exploiting local and topological information~\cite{ma2024efficient}. These methods mainly improve how MARL itself learns coordinated agent policies. In contrast, HRCL considers a different role for MARL. Instead of directly determining the detailed allocation of every agent, MARL learns higher-level guidance for an existing decentralized coordination process. 

\cparagraph{Hierarchical and Learning-Guided Multi-Agent Coordination}
Hierarchical reinforcement learning reduces complex decision problems by separating them into high- and low-level decisions~\cite{pateria2021hierarchical}. Temporal abstraction and action-space reduction have further been studied to reduce the number of decisions considered by reinforcement learning policies~\cite{machado2023temporal,majeed2021exact}. Similar ideas have been applied to cooperative multi-agent systems. Jendoubi and Bouffard~\cite{jendoubi2023multi} use high- and low-level policies for multi-agent energy management, while HAVEN~\cite{xu2023haven} introduces hierarchical coordination for cooperative MARL. ALMA~\cite{iqbal2022alma} learns high-level subtask allocation together with low-level agent policies. GoMARL~\cite{zang2023automatic} instead groups agents to improve cooperative learning efficiency. More recently, HCPO~\cite{liu2026hcpo} introduces a hierarchical conductor to guide policy optimization and coordinated exploration. These approaches demonstrate that hierarchy can reduce the difficulty of multi-agent learning, but their high- and low-level decisions are mainly represented by learned policies or learned coordination structures. HRCL uses hierarchy differently: its high level is learned by MARL, whereas its low level is an existing decentralized coordination algorithm. Reinforcement learning therefore guides rather than replaces decentralized coordination. The high-level policy controls the search space and the trade-off preference used by the lower layer, while the lower layer performs the detailed coordination.

\section{System Model}\label{sec:model}
In this section, the decentralized multi-agent coordination over a long-time span is modeled as a sequential decision-making. It involves the definition of plans, as well as modeling of individual and system-wide costs. 

\subsection{Problem Statement}
The core problem is \textit{how to coordinate multiple agents to select their operations (i.e., decision-making) among alternatives such that they complete system-wide tasks efficiently over a long time span (e.g., a full-day mission)}. 

Each option, referred to as a possible operational \emph{plan}, indicates a sequence of tasks scheduled over a specific time period for execution. This plan-based model is particularly powerful because each plan encapsulates a full operation cycle, allowing agents to reason at a higher level of abstraction. For example, a single plan can represent an extended drone operation, often lasting over 30 minutes, thereby eliminating the need for frequent minute-by-minute or second-by-second decision-making~\cite{ding2021crowdsourcing}. This significantly reduces the complexity of the decision space and allows the system to scale efficiently, even in large and dynamic environments.

For each agent $u$ and time period $t$, HRCL receives a finite, feasible, time-indexed candidate set $\mathcal{P}^u_t$ as an input loaded from the scenario data files. Each candidate plan $p^u_{kt} \in \mathcal{P}^u_t$ is a sequence of real values (i.e., a vector), formulated as follows:
\begin{equation}
    p^u_{kt} = (p^u_{ktd})_{d=1}^D \in \mathcal{P}^u_t,
    \label{eq:candidate_plan}
\end{equation}
where $d$ denotes the position index of a value inside a plan with the size of $D$; $k$ is the index of a candidate plan in $\mathcal{P}^u_t$, $k \leq K$. Equation~(\ref{eq:candidate_plan}) defines the supplied plan input and does not represent an online plan-generation operation.

Each agent selects one and only one plan $p^u_{s}$, $s \leq K$, which is referred to as the \textit{selected plan} (i.e. the agent's choice). All agents' choices aggregate element-wise to the collective choice, i.e., the \emph{global plan} $g$ of the multi-agent system, formulated as follows:
\begin{equation}
    g_{t} = (g_{td})_{d=1}^D = \sum_{u \in \mathcal{U}} p^u_{st}.
\end{equation}
Note that \emph{HRCL} assumes a finite feasible candidate plan set for each agent and addresses high-level plan group selection, behavior selection, and low-level decentralized selection among the supplied plans. Plan generation is domain-dependent and outside the proposed framework.

\subsection{Problem Formulation}
Based on the design of plans, the problem is defined by two objectives: 

One is to minimize the individual cost of agents, which is formulated by the \textit{discomfort cost function} $f_\text{d}(\cdot)$: It indicates the discomfort cost incurred by an agent executing a selected plan, e.g., the energy consumption of a drone flight path. The objective function for total discomfort cost denotes the time-accumulated discomfort cost of selected plans per agent over all time periods. It is formulated as follows:
\begin{equation}
    \mathop{\min}\limits_{p^u_{st} \in \mathcal{P}^u_t} \; \sum_{t \in \mathcal{T}} \sum_{u \in \mathcal{U}} f_\text{d}(p^u_{st}).
    \label{eq:objective1}
\end{equation}

The other objective is to minimize the system-wide cost, which is formulated by the \textit{inefficiency cost function} $f_\text{i}(\cdot)$: It aims to match the global plan of agents $g_t$ to the target of system $\tau_t$, e.g., reducing the Root Mean Square Error. The target $\tau_t$ is required for agents to complete at $t$, such as the required amount of data collected by drones. Since the cost function is non-linear, meaning the choices of the agents depend on each other, minimizing the inefficiency cost is a combinatorial NP-hard optimization problem~\cite{pournaras2018decentralized}. Therefore, the objective function of inefficiency cost is expressed as:
\begin{equation}
    \mathop{\min}\limits_{p^u_{st} \in \mathcal{P}^u_t} \; \sum_{t \in \mathcal{T}} f_\text{i}(\tau_t, g_t).
    \label{eq:objective2}
\end{equation}

Satisfying all of these (opposing) objectives depends on the selected plans by agents, e.g. whether they accept a plan with a bit higher discomfort cost to decrease inefficiency cost. Such multi-objective trade-offs are modeled as follows:
\begin{equation}
    \mathop{\min}\limits_{p^u_{st} \in \mathcal{P}^u_t} \; \sum_{t \in \mathcal{T}} \Big [ \sigma \cdot \sum_{u \in \mathcal{U}} f_\text{d}(p^u_{st}) + (1 - \sigma) \cdot f_\text{i}(\tau_t, g_t) \Big ],
    \label{eq:overall_objective}
\end{equation}
where $\sigma$ is the parameter set by the system to balance discomfort and inefficiency cost, $\sigma \in [0, 1]$.

\begin{figure}[!t]
    \centering
    \includegraphics[width=\linewidth]{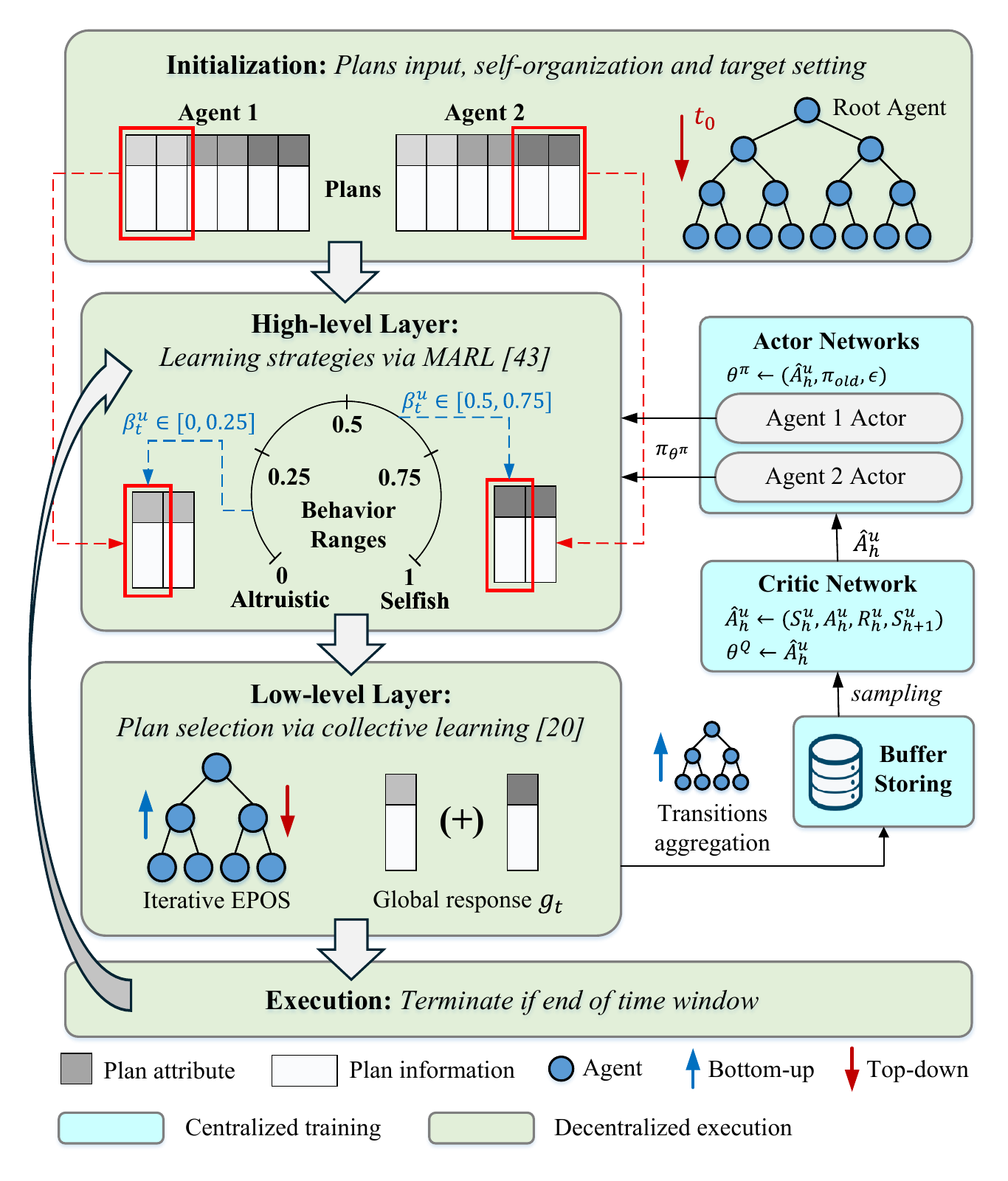}
    \caption{The system framework of \emph{HRCL}.}
    \label{fig:framework}
\end{figure}

\section{Proposed Method}\label{sec:proposed_method}
This section presents the proposed \emph{HRCL} framework. According to the overall objective in Eq.(\ref{eq:overall_objective}), the target $\tau_t$ updates over time, requiring agents to continuously adapt their decisions. To cope with this dynamic environments, agents strategically select current plans by predicting whether their collective long-term plans meet future targets. To this end, \emph{HRCL} leverages a MARL approach in which actors use compact local observations informed by aggregate collective outcomes and make autonomous decisions guided by long-term cumulative rewards aligned with the overall objective.

This formulation introduces two key challenges that \emph{HRCL} is designed to address: (1) While the proposed plan-based model mitigates the exponential growth of the action–state space by reasoning over long-term operations, they still require a large and diverse plan set $\mathcal{P}^u_t$ to approximate global optimality, causing the plan-selection problem to remain combinatorially large and challenging for MARL at scale. To mitigate this complexity, \emph{HRCL} decomposes decision-making into high-level action choices that capture both \emph{which alternatives agents consider} and \emph{what behaviors they adopt}. A behavior of agent $u$ at time period $t$, denoted as $\beta^u_t$, represents the agent's behavior on plan selection. As the value of $\beta^u_t$ increases, the agent becomes more selfish, prioritizing plans with low discomfort cost at the expense of higher inefficiency cost. Under these high-level actions, collective learning is used to coordinate agents’ plan selections. (2) Evaluating the overall reward requires aggregating the selected plans and associated discomfort costs across all agents, which can incur extremely high communication overhead. To reduce this cost, \emph{HRCL} adopts a tree-based communication topology for low-level plan aggregation and selection, where agents exchange plan information through structured bottom-up and top-down phases.

Fig.\ref{fig:framework} illustrates the overall framework of \emph{HRCL} based on a centralized training with decentralized high-level actors and tree-based coordination during execution. First, pre-generated, time-indexed candidate plan sets are loaded from scenario data files and supplied to each agent, such as load-shifting plans in energy self-management~\cite{pournaras2017self} and route plans in drone swarm sensing~\cite{qin2025strategic}. HRCL does not construct these candidate plans. Additionally, all agents are organized into a fixed balanced binary tree where the root agent receives the initial target task $\tau_0$ from the server and then sends it to other agents via the top-down phase. Next, at every time period $t$, each agent takes a high-level action, which is made with two types of strategies: \textit{grouping plans} and \textit{grouping behavior ranges}. The former strategy divides the currently applicable candidate plans into different groups based on their attributes, including discomfort cost, spatial direction and semantic information, and then chooses one of them. The latter strategy selects a behavior index corresponding to a representative value of $\beta^u_t$ to balance discomfort and inefficiency costs. Then, in the low-level layer, each agent coordinates to select its optimal plan based on the chosen behavior and obtains the global plan via collective learning. The global plan is used for reward calculation and state transition defined in Section~\ref{sec:rl_model}. Finally, agents execute their selected plans, whereas the system changes the environmental state and updates the target task according to $\tau_{t+1}=F(\tau_t,g_t)$; in the evaluated residual-target settings, $\tau_{t+1}=\tau_t-g_t$. During training, complete trajectories are stored directly in an on-policy rollout buffer at the central server; the EPOS tree is used only for low-level plan aggregation and selection.

\subsection{MARL modeling} \label{sec:rl_model}
At each time period, each actor receives a compact observation, makes a high-level decision, and obtains its next observation after collective plan selection and the environment transition. Thus, the problem can be modeled as a Markov decision process~\cite{bernstein2002complexity}. We model the problem using the concepts of observation, action and reward:

\cparagraph{Observation and state}
The local observation $o^u_t$ of agent $u$ is a compact local-and-aggregate summary produced by the environment and the collective learning outcome. It contains agent-specific and aggregated cost summaries returned after low-level collective plan selection, the raw target vector, and the complete global plan vector. The centralized critic uses the state $s_t = [o^1_t,\ldots,o^U_t]$, which concatenates all agents' local observations.

\cparagraph{Action}
The action of agent $u$ at $t$ is $ a^u_t=(i^u_t,m^u_t) \in \{1,\ldots,I\}\times\{1,\ldots,M\}$, where $i^u_t$ is the selected plan group index and $m^u_t$ is the selected behavior index (see Section~\ref{sec:method}). This pair can be encoded as one categorical action. After low-level plan selection, the environment transitions from $o^u_t$ to $o^u_{t+1}$.

\cparagraph{Reward}
The expected immediate reward is a shared team reward calculated from the normalized mean discomfort cost and normalized system-wide inefficiency cost. According to the overall objective function in Eq.(\ref{eq:overall_objective}), the reward function is expressed as:
    \begin{equation}
    r^u_t=r_t=-\sigma \cdot \bar{f}_{\text{d},t} -(1-\sigma) \cdot \bar{f}_{\text{i},t},
    \label{eq:reward}
    \end{equation}
where $\bar{f}_{\text{d},t}$ is the normalized mean discomfort cost and $\bar{f}_{\text{i},t}$ is the normalized system-wide inefficiency cost. The fixed weight $\sigma$ balances these system-level terms, whereas $\beta^u_t$ is an agent-specific low-level plan-selection preference selected by the high-level policy. The shared reward is appropriate because the system objective is fully cooperative.

\subsection{High-level Strategies}\label{sec:method}
The high-level actions provide strategic guidance and constraints, shaping the space of low-level actions to ensure that they align with overall objectives and efficiently achieve the optimization goals. The action consists of two strategies for group choices: the \emph{grouping plans} and the \emph{grouping behavior ranges}.

\cparagraph{Strategy for grouping plans} 
The key idea of this strategy is to divide all possible plans into multiple groups based on their associated attributes $\phi^u_{kt} \in \mathbb{R}^d$. The attribute could be (1) scalar discomfort cost of plans, denoted as $\phi^u_{kt} = f_\text{d}(p^u_{kt})$), (2) spatial direction (e.g., the flight range of drones that perform tasking), denoted as $\phi^u_{kt} = f_\text{s}(p^u_{kt})$, or (3) other semantic information. Since each agent selects a group of plans instead of choosing directly from the entire set $\mathcal{P}^u_t$, the number of available actions is significantly reduced, leading to a smaller action space. The grouping function is defined as follows:
\begin{equation}
    \mathcal{G} : \{ \phi^u_{kt} \}^K_{k=1} \rightarrow \{ 1,..,I \},
\end{equation}
\begin{equation}
    \mathcal{P}^u_{it} \triangleq \Big\{ p^u_{kt} \; | \; \mathcal{G}(\phi^u_{kt}) = i \Big \} = \{ p^u_{ijt} \}^{P_i}_{j=1},
\end{equation}
where $i$ denotes the index of a plan group within a total of $I$ groups; $j$ implies the index of a plan within group $i$; $P_i$ is the total number of plans within group $i$, $P_i \triangleq |\mathcal{P}^u_{it}| \leq K$; $\mathcal{G}(\phi^u_{kt}) = i$ assigns plan $p^u_{kt}$ to group $i$ based on its attribute $\phi^u_{kt}$. After that, each agent $u$ takes action $i^u_t$ to select a group of plans $\mathcal{I}^u_t$ from all possible plan groups $\mathcal{P}^u_{it}$, $\forall i \leq I$, which is formulated as:
\begin{equation}
    \mathcal{I}^u_t = \mathcal{P}^u_{it} \; | \; i = i^u_t.
    \label{eq:group_plan_select}
\end{equation}

\cparagraph{Strategy for grouping behavior ranges} 
This strategy is designed to select from a group of behavior range rather than choosing among all possible values. The entire range $[0,1]$ splits into $M$ non-overlapping ranges of equal length, each range with an index of $m$. Then, each agent takes the action $m^u_t$ to choose a range, and sets the mean value of the range as the agent's behavior $\beta^u_t$, $0 \leq \beta^u_t \leq 1$. The strategy can be formulated as:
\begin{equation} 
    \beta^u_t = \frac{m}{M} - \frac{1}{2M} \; | \; m = m^u_t.
    \label{eq:behavior}
\end{equation}

\begin{figure}[!t]
    \centering
    \includegraphics[width=\linewidth]{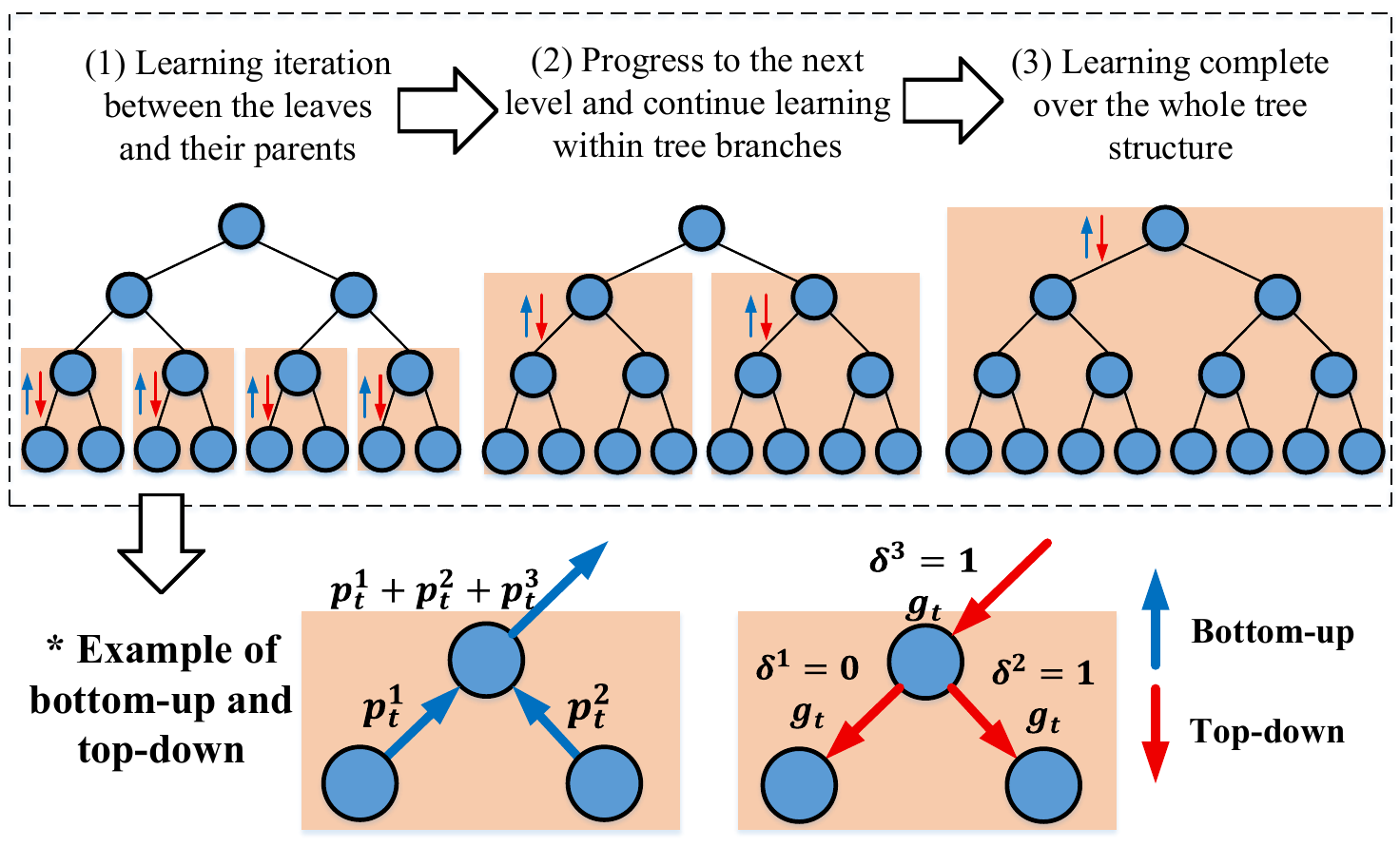}
    \caption{The learning iteration in the low-level plan selection. During the bottom-up phase, Agent 3 aggregates the plans of its children, i.e., Agent 1 and Agent 2, and sends them to its parent agent with its own plan. During the top-down phase, each parent agent sends the global plan and a decision to approve/reject the plan selection to its children.}
    \label{fig:plan_select}
\end{figure}

\subsection{Low-level Plan Selection} \label{sec:collective_learning}
With the input of selected groups of plans $\mathcal{I}^u_t$ and behaviors $\beta^u_t$ through the high-level strategies, each agent selects an optimal plan using collective learning~\cite{pournaras2018decentralized}. This method uses a fixed balanced binary tree to structure the agents' interactions for each low-level optimization call in the current study. Agents connect into the tree structure within which they iteratively interact with their children and parent, i.e, the neighboring agents, in order to observe and aggregate the plans efficiently to perform decisions. The collective-learning preliminaries are summarized in Appendix~\ref{app:collective_learning}. 

As shown in Fig.~\ref{fig:plan_select}, the communication process at each iteration contains two phases: bottom-up and top-down. During the bottom-up phase, each agent (except for leaf nodes) aggregates the plans of its children and selects its own plan via:
\begin{equation}
    \mathop{\min}\limits_{p^u_{st} \in \mathcal{I}^u_t} \; \beta^u_t \cdot f_\text{d}(p^u_{st}) + (1 - \beta^u_t) \cdot f_\text{i}(\tau_t, g_t).
    \label{eq:epos}
\end{equation}
Then, the agent combines the selected plan with the aggregated plans from its children and shares this aggregated information to its parent node (if it is not the root). As a result, the root node aggregates the plans of all agents and obtains the global plan. During the top-down process, however, each agent (except for leaf nodes) sends feedback to its children, including the global plan $g_t$ and a decision $\delta^u$ regarding whether to approve the plan selection of its children in the current iteration. The decision is calculated by comparing the inefficiency cost based on the selected plan of child agent in the current iteration with the outcome in the previous iteration~\cite{pournaras2018decentralized}. An agent node will adopt the new selected plan if approved by its parent ($\delta^u = 1$), i.e., the current inefficiency cost is lower than the previous one. Otherwise, it keeps the plan selected in the previous iteration. Finally, all agents store the global plan for the next iteration.


\begin{algorithm}[!t]
    \caption{\emph{HRCL} Training.}
    \small
    \label{algorithm1}
    Randomly initialize value critic $V_\psi(\cdot)$ and actor $\pi_\theta(\cdot)$\;
    \For{rollout batch $:= 1$ to max-rollout-number}
    {
        Initialize the on-policy rollout buffer\;
        \For{each complete episode in the rollout batch}
        {
            Reset the target and obtain $o^u_1$ and $s_1=[o^1_1,\ldots,o^U_1]$\;
            \For{period $t := 1$ to max-episode-length}
            {
                Retrieve the pre-generated candidate plan sets $\{\mathcal{P}^u_t\}_{u\in\mathcal{U}}$ for the current period\;
                \For{$\forall u \in \mathcal{U}$}
                {
                    Select $a^u_t=(i^u_t,m^u_t) \sim \pi_\theta(\cdot|o^u_t)$\;
                    Choose $\mathcal{I}^u_t$ and $\beta^u_t$ via Eqs.(\ref{eq:group_plan_select})--(\ref{eq:behavior})\;
                }
                Coordinate low-level plan selection via collective learning\;
                Obtain $o^{1:U}_{t+1}$, $s_{t+1}$, shared $r_t$, and terminal indicators $d^{1:U}_t$\;
                Store $o^{1:U}_t,s_t,a^{1:U}_t$, old log probabilities, $r_t$, $V_\psi(s_t)$, and $d^{1:U}_t$\;
            }
        }
        Compute GAE via Eq.(\ref{eq:advantage})\;
        Update $\theta$ and $\psi$ using rollout mini-batches via Eqs.(\ref{eq:prob})--(\ref{eq:critic})\;
        Clear the rollout buffer\;
    }
\end{algorithm}

\begin{algorithm}[!t]
    \caption{\emph{HRCL} Execution.}
    \small
    \label{algorithm2}
    \textbf{Input:} Agent $u$, $\tau_0$, $\pi_\theta(\cdot)$ \;
    \textbf{Output:} Selected plans $(p^u_{s1},...,p^u_{sT})$, \;
    \For{period $t := 1$ to max-episode-length}
    {
        Update the target $\tau_t$ and obtain the local observation $o^u_t$\;
        Retrieve the pre-generated candidate plan set $\mathcal{P}^u_t$ for the current period\;
        Take action: $a^u_t=(i^u_t,m^u_t) \sim \pi_\theta(\cdot|o^u_t)$\;
        Grouping plans and choose $\mathcal{I}^u_t$ via Eq.(\ref{eq:group_plan_select})\;
        Choose the representative behavior $\beta^u_t$ via Eq.(\ref{eq:behavior})\;
        \For{iteration $:= 1$ to max-iteration-number}
        {
            Aggregate the plans of children and determine decision $\delta^u$\;
            Select a plan $p^u_{st}$ via Eq.(\ref{eq:epos}) \;
            Share the aggregated plans and selected plan to parent\;
            Wait for the decision and the global plan $g_t$ from parent\;
            Send $\delta^u$ and $g_t$ to children\;
        }
        
    }
\end{algorithm}

\subsection{MARL Training and Execution} \label{sec:training}
The objective of training the \textit{HRCL} framework is to optimize the parameters of deep neural networks (DNNs) representing agents’ policies so they can reliably achieve high rewards across multiple episodes. Each episode reflects a complete cycle of plan selection and decision-making over a temporal horizon. Once the training process is complete, the learned policies are deployed as the execution mechanism for \textit{HRCL}.

During centralized training, the state-value critic receives $s_t=[o^1_t,\ldots,o^U_t]$, while each actor uses only its local observation through $\pi_\theta(a^u_t|o^u_t)$. A batch of complete episodes is collected in an on-policy rollout buffer, including joint observations, centralized states, actions, old log probabilities, shared rewards, value estimates, and terminal indicators. GAE is then calculated, PPO updates are performed using mini-batches, and the rollout buffer is cleared. During decentralized execution, each high-level actor selects its action from its own observation, while the low-level plan-selection process remains decentralized through tree-based message passing. The tree does not aggregate state--action--reward transitions into the rollout buffer.

To enhance learning performance, \textit{HRCL} utilizes an actor-critic architecture that combines the strengths of policy-based and value-based methods~\cite{lowe2017multi}. It uses (1) a centralized state-value critic $V_\psi(s_t)$ at the central server and (2) a local actor $\pi_\theta(a^u_t|o^u_t)$ for each agent. This setup allows the actor to learn directly from gradient feedback, while the critic provides stabilizing state-value estimates. To further reinforce stability and prevent overly aggressive policy updates, \textit{HRCL} incorporates Proximal Policy Optimization (PPO)~\cite{nayak2023scalable}, a widely used method for regularizing policy changes.

During training, the temporal-difference residual and generalized advantage estimate (GAE) are calculated over each complete trajectory:
\begin{align}
    \delta^u_t={}&r^u_t+\gamma\cdot(1-d^u_t)\cdot V_\psi(s_{t+1})-V_\psi(s_t), \notag\\
    \hat{A}^u_t={}&\sum_{l=0}^{T-t-1}(\gamma\cdot\lambda)^l\cdot\delta^u_{t+l},
    \label{eq:advantage}
\end{align}
where $\gamma \in [0,1]$ is the discount factor, $\lambda \in [0,1]$ is the GAE parameter, and $d^u_t$ is the terminal indicator. The advantage $\hat{A}^u_t$ is passed to the actor network to encourage beneficial actions and discourage less effective ones. PPO stabilizes learning by constraining the local policy update ratio:
\begin{equation}
   \rho^u_t(\theta)=
   \frac{\pi_\theta(a^u_t|o^u_t)}{\pi_{\theta_{\mathrm{old}}}(a^u_t|o^u_t)},
   \label{eq:prob}
\end{equation}
where $\pi_{\theta_{\mathrm{old}}}$ denotes the actor policy before the current PPO update. This ratio is used in the clipped surrogate term:
\begin{align}
    \begin{split}
    \hat{C}^u_t(\theta)=\min\Big[{}
    &\rho^u_t(\theta)\cdot\hat{A}^u_t, \\
    &\operatorname{clip}\big(\rho^u_t(\theta),1-\epsilon,1+\epsilon\big)
     \cdot\hat{A}^u_t\Big],
    \label{eq:clip_obj}
    \end{split}
\end{align}
with the hyperparameter $\epsilon$ serving as a clipping threshold to limit how far the policy is allowed to change. The actor loss includes the required negative sign for minimization and an entropy term used to encourage exploration:
\begin{equation}
    \begin{split}
    L_{\text{actor}}(\theta)={}&-\frac{1}{|\mathcal{B}|}
    \sum_{(u,t)\in\mathcal{B}}\hat{C}^u_t(\theta) \\
    &-c_{\mathrm{ent}}\cdot\frac{1}{|\mathcal{B}|}
    \sum_{(u,t)\in\mathcal{B}}\mathcal{H}\big[\pi_\theta(\cdot|o^u_t)\big],
    \end{split}
    \label{eq:actor}
\end{equation}
The critic target uses the value estimate stored before the PPO update, and the critic minimizes its mean squared error:
\begin{align}
    V^u_{\mathrm{target},t}={}&\hat{A}^u_t+V_{\mathrm{old}}(s_t), \notag\\
    L_{\text{critic}}(\psi)={}&\frac{1}{|\mathcal{B}|}
    \sum_{(u,t)\in\mathcal{B}}
    \big(V_\psi(s_t)-V^u_{\mathrm{target},t}\big)^2,
    \label{eq:critic}
\end{align}

In summary, the training process primarily involves the following steps (see Algorithm~\ref{algorithm1}): It first initializes the actor and value critic. It then collects a rollout batch of complete episodes in which actors choose actions from local observations and collective learning performs low-level plan selection. After the rollout batch is complete, GAE is calculated and the actor and value critic are updated using mini-batches. Finally, the on-policy rollout buffer is cleared before collecting the next batch.

Moreover, the decentralized execution process is illustrated in Algorithm~\ref{algorithm2}. With the input of initial target $\tau_0$, local policy $\pi_\theta(\cdot)$ and high-level strategy, each agent runs the algorithm and outputs $p^u_{st}$ at each time period. Each high-level actor uses only $o^u_t$, while low-level plan selection exchanges aggregated plans and feedback through the fixed EPOS tree. Finally, each agent obtains its plans for all periods and executes them.

\section{Experimental Methodology}\label{sec:ex_method}
This section defines the hyperparameter of the proposed approach. The metrics and baselines used for the performance evaluation are introduced. The coordination and plan-selection computation and communication complexity of all methods after candidate plans are available are also compared.


\subsection{Scenario Settings}
The synthetic scenario is built by extracting data from the synthetic dataset~\cite{Pournaras2023}. It contains $1,000$ agents, each with $16$ possible plans (i.e., vectors) that consists of a sequence of $100$ values sampled from a Normal distribution centered at a mean of 0 with a variance of 1. The discomfort costs of plans for each agent are increasing linearly within the range $[0, 1]$. The plans are grouped using the quantile-based discretization function based on their discomfort cost, i.e., the attribute function is $\phi^u_{kt} = f_\text{d}(p^u_{kt})$. 

The goal of the optimization is to bring the global plan close to a target signal (a cosine waveform) over $16$ time periods, while minimizing the discomfort cost of all agents. At each time period $t$, the update of the target can be formulated as: $\tau_{t+1} = \tau_{t} - g_t$, where $\tau_0$ represents the target signal of the cosine waveform. Therefore, each current allocation changes the residual task that must be completed through subsequent coordination decisions. We also use $80\%$ plans of the datasets for training and $20\%$ for testing. This makes the global plan differentiated in each test and the task environment dynamically changing.

There are two types of scenarios: (1) \emph{Basic synthetic scenario.} It is a benchmark with $40$ agents, $16$ time periods, $32$ plans per agent and the target signal of $\omega = \pi / 24$, aiming to compare the performance of different approaches. (2) \emph{Complex synthetic scenarios.} They are studied by varying the parameters to validate the scalability of the proposed solution. There are three dimensions: (1) number of agents, (2) number of plans (i.e., the plan volume owned by each agent), and (3) complexity of target tasks (i.e., different frequency multipliers of cosine waveforms). 

Apart from synthetic scenarios, this paper studies two smart city application scenarios: energy self-management (Section~\ref{sec:energy_manage}) and drone swarm sensing (Section~\ref{sec:drone_sensing}). Table~\ref{tab:dataset} shows the comparison of parameter settings among different scenarios. In the synthetic, energy, and drone experiments, application-specific procedures construct the candidate plans upstream and store the pre-generated, time-indexed plan sets in scenario data files. HRCL loads these sets as inputs. Within each experiment, all compared methods receive the same candidate plans.

\begin{table*}[!t]
\centering
\caption{Parameters for datasets.}
\label{tab:dataset}
 \begin{tabular}{l|p{4cm}|p{4.5cm}|p{4cm}}
    \toprule
    Parameter	&	Synthetic 	&	Energy  & Drone   \\ \hline

    System-wide goals
    & \tabincell{l}{Cosine waveform shaping} & \tabincell{l}{Power peak prevention} & \tabincell{l}{Traffic vehicle detection} \\

    Grouping plans
    & $\phi^u_{kt} = f_\text{d}(p^u_{kt})$ & $\phi^u_{kt} = f_\text{d}(p^u_{kt})$ & $\phi^u_{kt} = f_\text{s}(p^u_{kt})$\\
    
    No. of agents ($U$)
    & 40 / 80 / 120 / 160 / 200 & 160 & 16 \\ 
    
    No. of plans ($K$)
    & 32 / 64 / 96 / 128 & 10 & 64\\
    
    No. of time periods ($T$)
    & 16 & 16 & 16 \\

    Candidate-plan procedure
    & Normal distribution~\cite{pournaras2018decentralized} & Load-shifting~\cite{pournaras2017self} & Route planning~\cite{qin2025strategic}\\
    
    Inefficiency cost
    & Mismatch between global plan and target signal  & Load imbalance of smart grids & Mismatch between drone sensing and requirements \\

    Discomfort cost
    & Plan index & Minutes shifted from original demand & Energy consumption of drones\\
    
    \bottomrule
    \end{tabular}
\end{table*}

\subsection{Metrics and Baselines}\label{sec:metric_base}
To evaluate the performance of \emph{HRCL}, we introduce three metrics according to Eq.(\ref{eq:reward}): (1) \textit{Mean discomfort cost} $\bar{f}_{\text{d},t}$, which denotes the average of discomfort cost of all agents; (2) \textit{Inefficiency cost} $\bar{f}_{\text{i},t}$, which calculates the Root Mean Square Error (RMSE) between the target and global plan; and (3) \textit{Combined cost}, which measures the weighted sum of the normalized mean discomfort cost and inefficiency cost. 

A fair comparison of the proposed methods with related work is not straightforward, as there is a very limited number of relevant decentralized algorithms. Similar algorithms, such as ALMA, GoMARL and HCPO~\cite{iqbal2022alma,zang2023automatic,liu2026hcpo}, cannot be directly applicable to the plan-based optimization problem defined in Section~\ref{sec:model}. For this reason, this paper focuses on ablation studies by testing standalone reinforcement learning and collective learning. The baseline methods are introduced as: (1) \emph{EPOS}: This collective learning~\cite{majumdar2023discrete} is adapted to coordinate agents to select a plan from the set $\mathcal{P}^u_t$ via Eq.(\ref{eq:epos}) rather than $\mathcal{I}^u_t$ without any long-term strategic decision on task planning via learning methods. (2) \emph{MAPPO}: It is a state-of-the-art MARL technique~\cite{nayak2023scalable} adapted where each agent takes an action to choose one plan from the set $\mathcal{P}^u_t$. It employs the reward function of Eq.(\ref{eq:reward}) without high-level strategies for group choices. (3) \emph{HRL}: It is adapted from hierarchical RL~\cite{jendoubi2023multi} that each agent takes an action for high-level choices of plan groups and then takes a second action to select a plan within the chosen group (i.e., low-level plan selection). The policy networks of both types of actions are trained independently by using two different actor networks. (4) \emph{OR-Tools}: This centralized baseline uses CP-SAT~\cite{cpsatlp} to aggregate the $P^u_t$ from all agents and return the lowest combined cost result within the 120-minute timeout. This method only addresses low-level plan selection without high-level strategies. (5) \emph{Greedy}: Agents only choose their plans with the minimum discomfort cost.

Furthermore, two variants of the proposed \emph{HRCL} are considered: (1) \emph{HRCL-P}, which uses the strategy for \emph{grouping plans} only and sets the agent behaviors the same ($\beta^u_t=0.5$), and (2) \emph{HRCL-B}, which uses the strategy for \emph{grouping behavior ranges} only where agents select plans from the whole set $\mathcal{P}^u_t$. As a comparison, \emph{HRCL} takes both strategies together.

\subsection{Algorithm Settings}\label{sec:algorithm_set}
The setting of the proposed approach\footnote{HRCL is available at: https://github.com/TDI-Lab/HRCL.} contains both collective learning and reinforcement learning parts: During the coordinated plan selection via EPOS\footnote{EPOS is available at: https://github.com/epournaras/EPOS.}, the communication structure among agents is set as a fixed balanced binary tree~\cite{majumdar2023discrete}. In the one execution of EPOS, the agents perform $20$ bottom-up and top-down learning iterations. For the reinforcement learning in \emph{HRCL}, complete episodes are collected in an on-policy rollout buffer. After GAE is calculated, PPO updates use mini-batches of size $H=64$, and the rollout buffer is then cleared. The discount factor is $\gamma=0.95$, and the clip interval hyperparameter is $0.2$. We use the recurrent neural network (RNN), and use $W=64$ neurons in the two hidden layers of the RNN in both critic and actor networks. The activation function used for the networks is tanh. The models are trained over $\mathcal{E}=2000$ episodes, each consisting of multiple time periods. Furthermore, our approach employs Proximal Policy Optimization to prevent detrimental updates and improving the stability of the learning process~\cite{nayak2023scalable}.

In addition, several parameters in the proposed methods are empirically selected due to the optimal performance: \emph{HRCL-P} divides plans into $4$ groups, each containing $4$ plans; \emph{HRCL-B} divides the behavior range into $4$ ranges; and \emph{HRCL} groups both $4$ plans and $2$ behavior ranges. The fixed weight $\sigma$ in Eq.(\ref{eq:reward}) is set as $0.5$ such that the reward is equivalent to the metric of combined cost defined in Section~\ref{sec:metric_base}. Moreover, the agent behaviors $\beta^u_t$ in baseline \emph{EPOS} are assumed to be the same, and empirically selected according to the lowest selected combined cost.

\section{Performance Evaluation}\label{sec:performance}
In this section, the synthetic scenario is used to validate the superior performance of the proposed \emph{HRCL}. Additionally, this section shows the broad and significant impact of the proposed algorithm on two very different scenarios in the context of smart grids and smart cities: (1) Energy self-management and (2) drone swarm sensing. 

\begin{table*}[!t]
    \centering
    \scriptsize
    \caption{Cost comparison of all methods in the basic synthetic scenario.}
    \label{table:basic}
    \resizebox{0.7\linewidth}{!}{
    \begin{tabular}{lccccccccc}
        \toprule
        & \multicolumn{3}{c}{\textbf{Inefficiency Cost}} & \multicolumn{3}{c}{\textbf{Mean Discomfort Cost}} & \multicolumn{3}{c}{\textbf{Combined Cost}} \\
        \cmidrule(lr){2-4} \cmidrule(lr){5-7} \cmidrule(lr){8-10}
        \textbf{Method} & Mean & Err & Gap & Mean & Err & Gap & Mean & Err & Gap \\
        \midrule
        
        \emph{OR-Tools}        & \cellcolor{blue!10}0.213 & 0.0 & 0.0\%    & \cellcolor{blue!10} 0.252 & 0.0 & 0.0\%     &\cellcolor{blue!10} 0.232 & 0.0 & 0.0\% \\
        \emph{Greedy}    & 0.932 & 0.011 & 337.5\%  & 0.0 & 0.0 & -100.0\%  & 0.466 & 0.011 & 100.6\% \\
        \emph{EPOS}          & 0.259 & 0.025 & 21.78\%   & 0.272 & 0.010 & 7.99\%     & 0.266 & 0.034 & 14.29\% \\
        \emph{MAPPO}           & 0.311 & 0.007 & 45.87\%   & 0.236 & 0.003 & -6.36\%    & 0.273 & 0.010 & 17.56\% \\
        \emph{HRL}             & 0.303 & 0.007 & 42.25\%   & 0.195 & 0.003 & -22.62\%    & 0.249 & 0.010 & 7.33\% \\
        \emph{HRCL-P}          & \cellcolor{yellow!30}0.244 & 0.013 & 14.37\%   & \cellcolor{yellow!30}0.162 & 0.003 & -35.75\%   & \cellcolor{yellow!30}0.203 & 0.016 & -12.79\% \\
        \emph{HRCL-B}          & \cellcolor{yellow!30}0.181 & 0.016 & -14.84\%  & \cellcolor{yellow!30}0.243 & 0.007 & -3.26\%    & \cellcolor{yellow!30}0.212 & 0.023 & -8.59\% \\
        \emph{HRCL}            & \cellcolor{yellow!30}0.200 & 0.016 & -5.96\%   & \cellcolor{yellow!30}0.172 & 0.003 & -31.69\%   & \cellcolor{yellow!30}\textbf{0.186} & 0.019 & \textbf{-19.91\%} \\
        \bottomrule
    \end{tabular}}
\parbox{0.85\linewidth}{\small \textit{Note.} The setting of the basic scenario is $40$ agents, $32$ plans per agent, and a target signal with $\omega=\pi/24$). Gap is the percentage difference between each mean and the corresponding \emph{OR-Tools} mean; negative values indicate lower costs.}
\end{table*}

\subsection{Evaluation on basic synthetic scenario} \label{sec:eval_synthetic_results}
Table~\ref{table:basic} illustrates the cost advantages of the proposed approach. \emph{HRCL-P} achieves a combined cost that is $23.69\%$ lower than \emph{EPOS}, benefiting from the use of DNNs for function approximation. This assists \emph{HRCL-P} to observe the environment and strategically select plan groups that reduce discomfort cost over time, while maintaining low inefficiency cost. Similarly, \emph{HRCL-B} optimizes behavior ranges for agents by recommending them to behave selfishly at the beginning and then altruistically, minimizing combined costs. However, \emph{HRCL-B} tends to coordinate agents toward plans with higher discomfort cost, leading to a combined cost that is $4.81\%$ higher than \emph{HRCL-P}. By integrating the strengths of both high-level strategies, \emph{HRCL} learns to optimize both plan groups and agent behaviors, achieving the lowest combined cost, which is $19.91\%$ lower than the best solution returned by OR-Tools.

With the help of effective coordination through collective learning, \emph{HRCL} avoids unnecessary exploration, allowing it to more efficiently converge on the optimized global plan. As a result, it achieves $35.53\%$ lower inefficiency cost and $27.05\%$ lower discomfort cost compared to \emph{MAPPO}. Note that the low-level policy on plan selection of \emph{HRL} heavily depends on its high-level policy on plan group selection. The changing high-level policy in \emph{HRL} leads to the plans selected by low-level policy are changing even taking the same action, resulting in non-stationary training and $27.79\%$ higher combined cost than \emph{MAPPO}. 

\begin{figure*}[!t]
    \centering
    \subfigure[The training process.]{
		\includegraphics[width=0.22\linewidth]{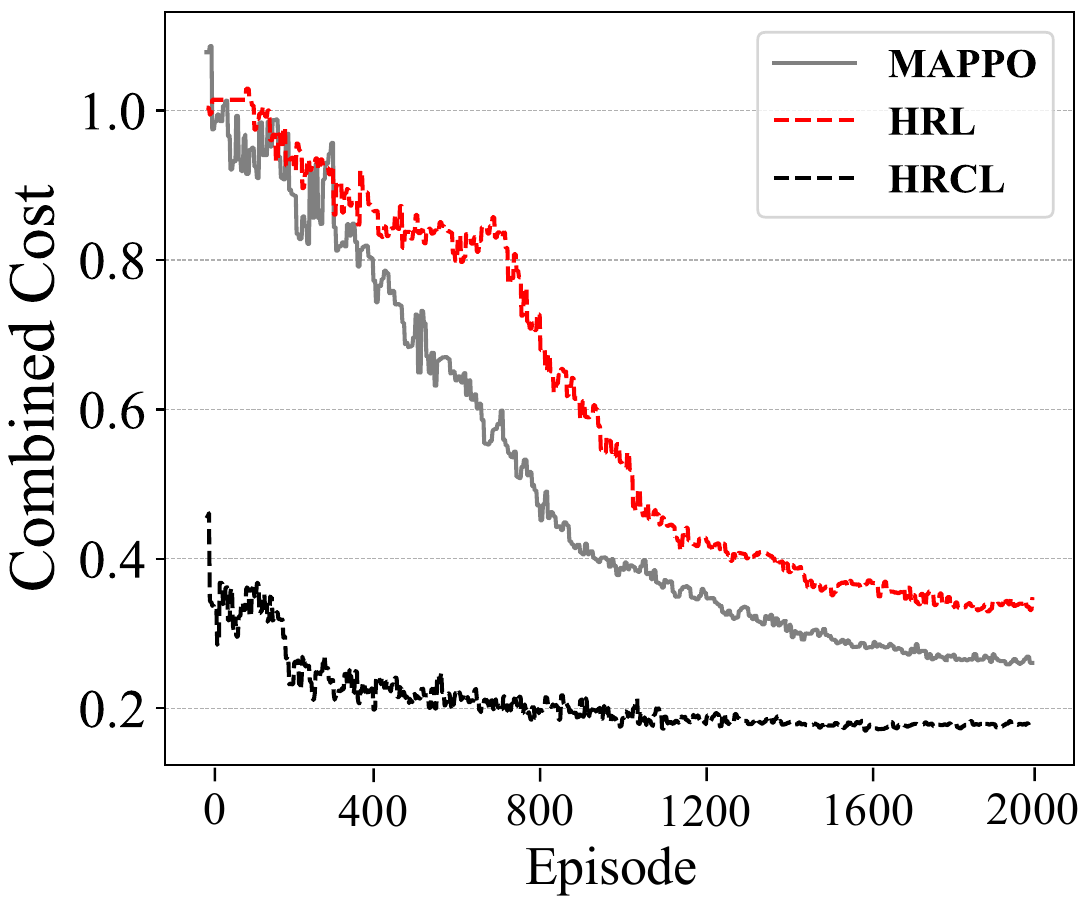}
        \label{fig:training}
	}
    \subfigure[Cost per time period.]{
		\includegraphics[width=0.22\linewidth]{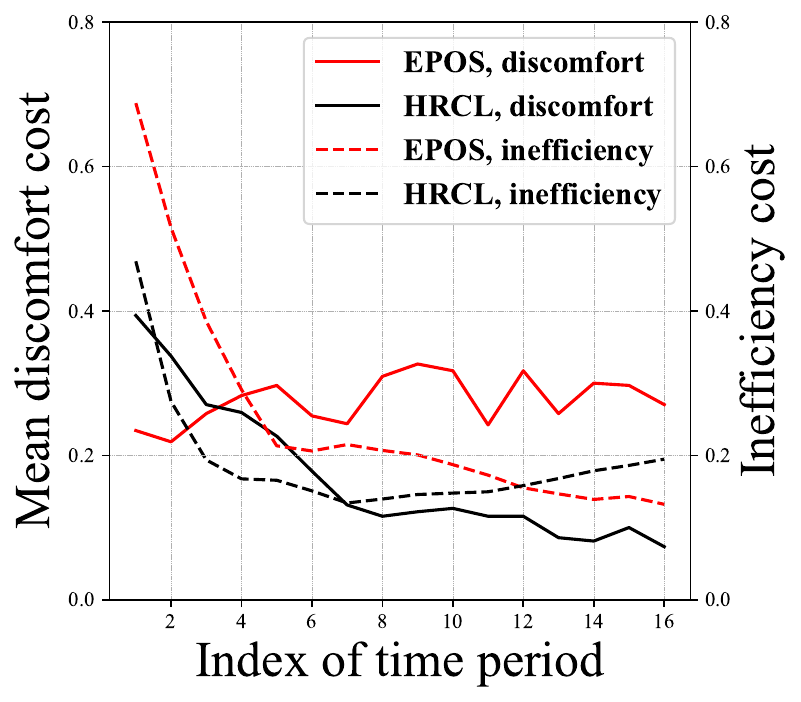}
        \label{fig:cost_time}
	}
    \subfigure[Different weight values.]{
		\includegraphics[width=0.22\linewidth]{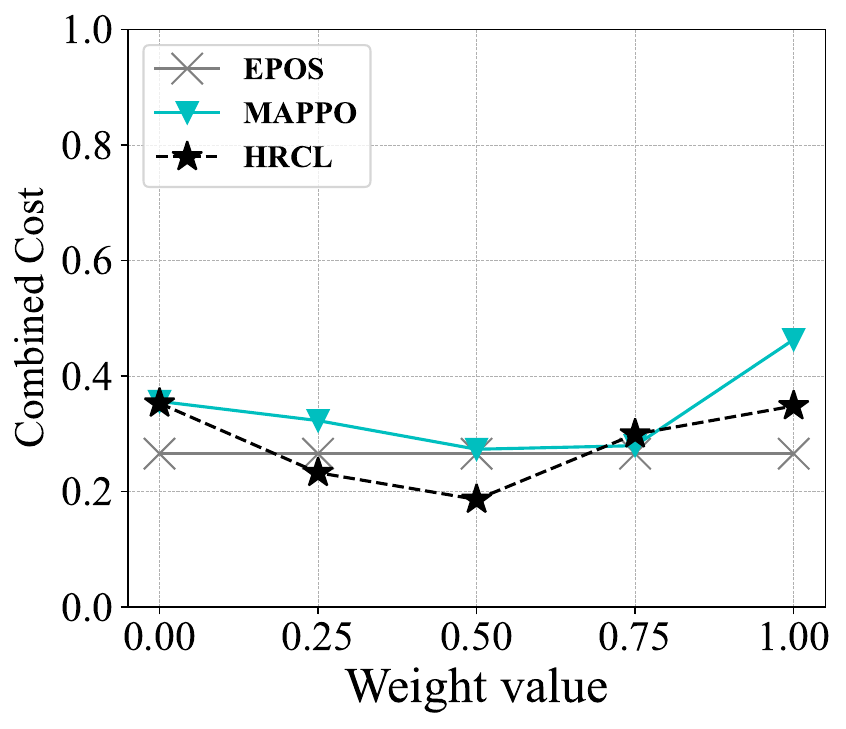}
        \label{fig:weight}
	}
    \subfigure[Different behavior values.]{
		\includegraphics[width=0.22\linewidth]{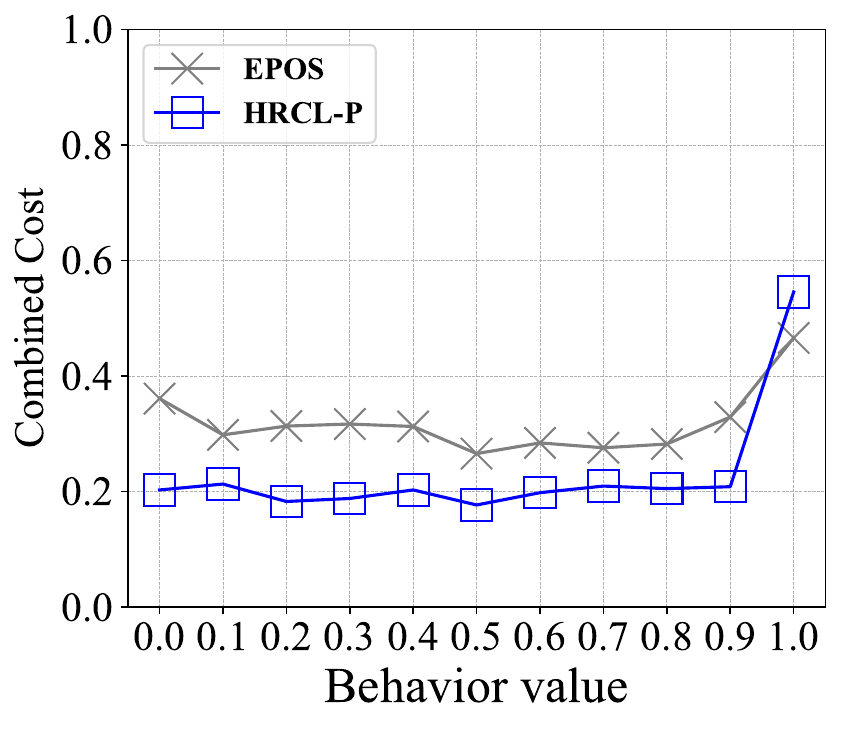}
        \label{fig:behavior}
	}
    \subfigure[Different actions.]{
		\includegraphics[width=0.23\linewidth]{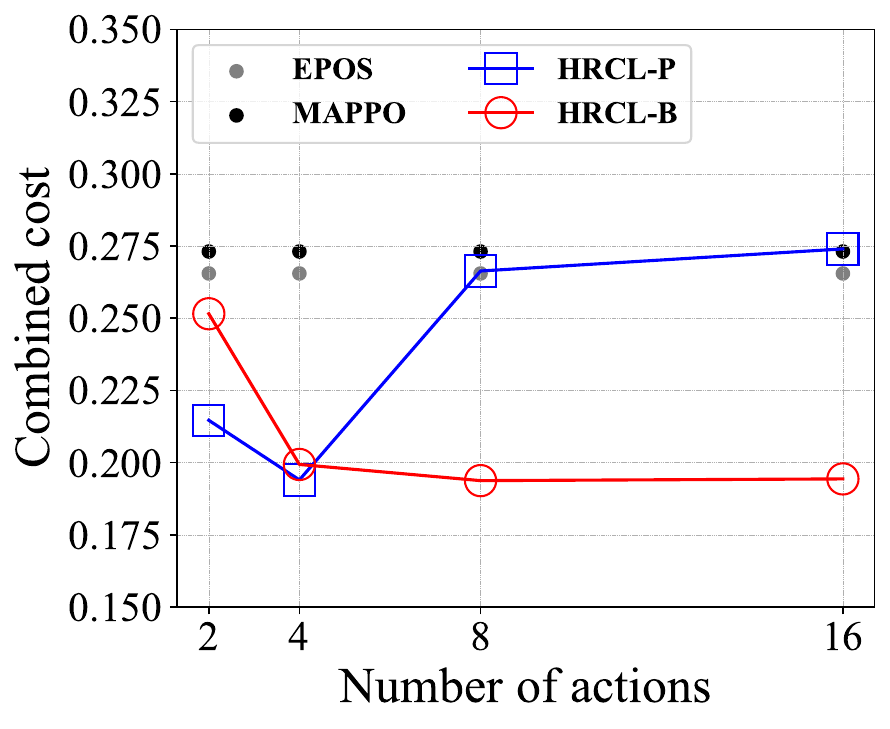}
        \label{fig:actions}
	}
    \subfigure[Average frequency vs. discomfort cost of groups.]{
		\includegraphics[width=0.32\linewidth]{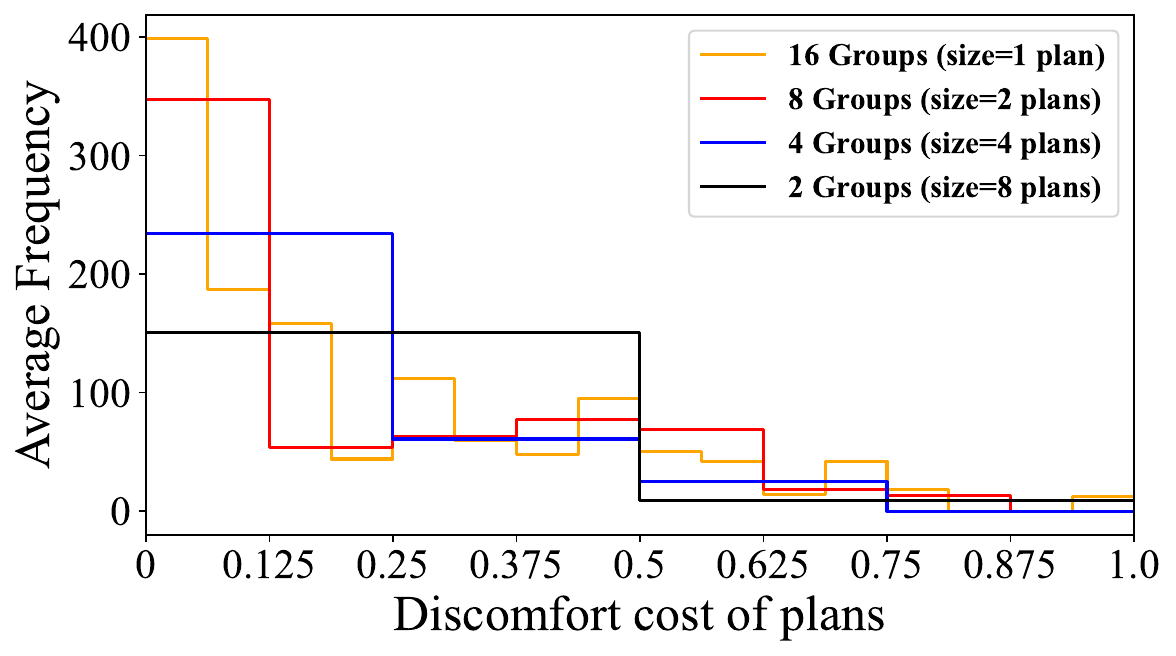}
        \label{fig:frequency_group}
	}
    \subfigure[Average frequency vs. behavior range.]{
		\includegraphics[width=0.32\linewidth]{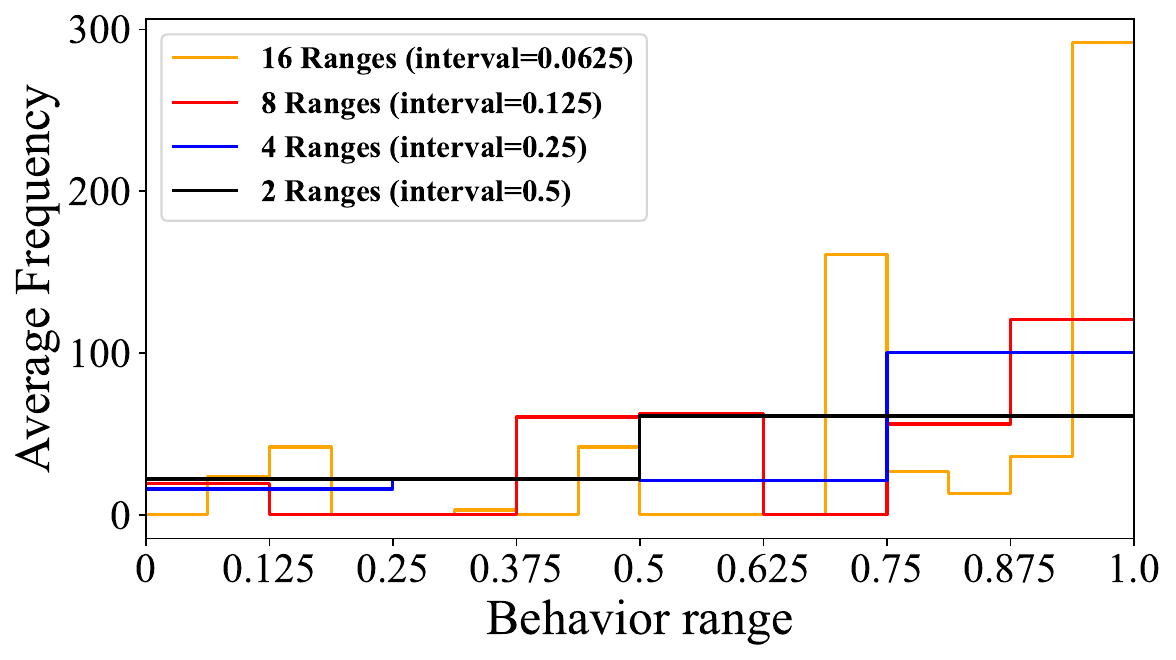}
        \label{fig:frequency_behavior}
	}
    \caption{Comparison of training convergence, performance over time, and different parameters, including weights in reward function, agents' behavior, and the number of actions.}
    \label{fig:parameters}
\end{figure*}

Fig.~\ref{fig:training} shows the training process of four MARL-based methods. With the help of low-level collective learning, \emph{HRCL} achieves faster exploration and converges around $600$ episode, significantly earlier than \emph{MAPPO}, which converges after $2000$ episode. Fig.~\ref{fig:cost_time} compares the cost performance of different methods over time. The proposed \emph{HRCL} shows consistent improvement and eventually surpasses \emph{EPOS}. Specifically, by time period $4$, \emph{HRCL} achieves a lower mean discomfort cost than \emph{EPOS}, and although its inefficiency cost exceeds that of \emph{EPOS} at time period $13$, its discomfort cost remains $66.31\%$ lower, resulting in a lower combined cost. These findings highlight the long-term benefits of high-level strategy learning in MARL, particularly over extended time horizons.

\begin{figure}
    \centering
    \includegraphics[width=\linewidth]{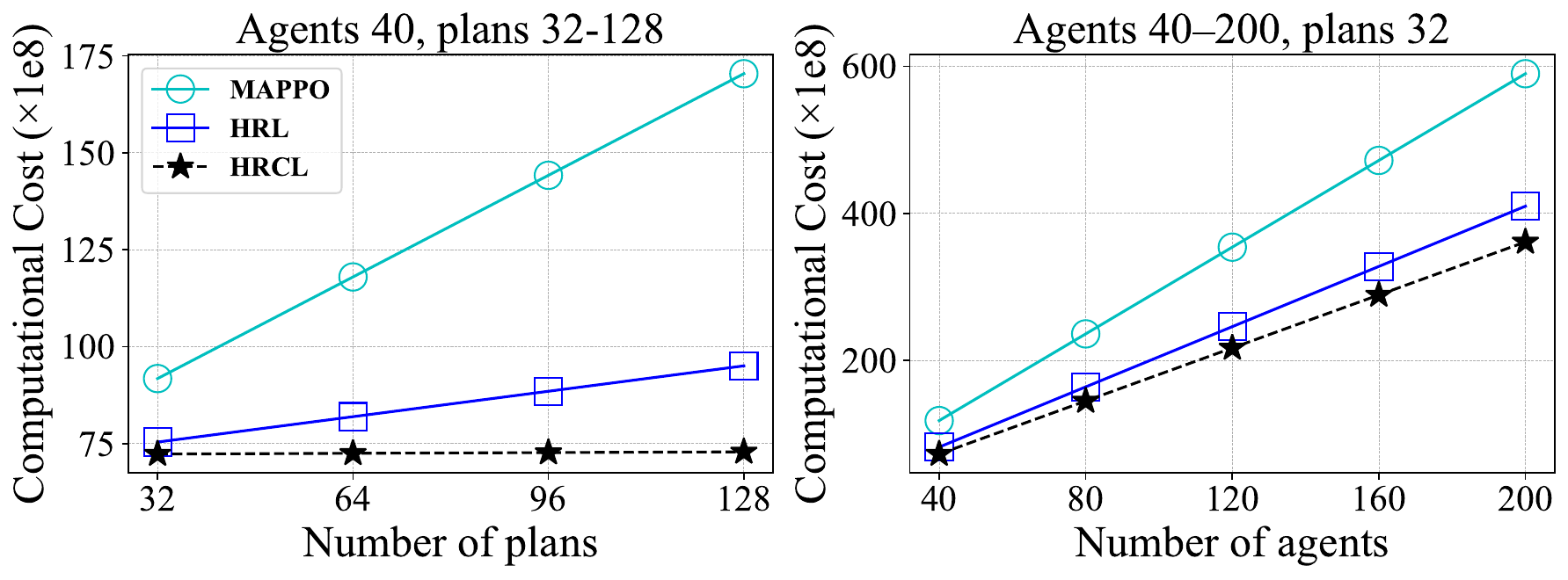}
    \caption{Comparison of computational overhead over under different number of plans and agents.}
    \label{fig:complexity}
\end{figure}

\subsection{Complexity Analysis}
We further analyze the computation required for learning and coordination after candidate plans are available. Candidate-plan construction is excluded because it is an upstream, application-specific process shared by all compared methods. Therefore, the following analysis focuses only on policy inference, plan grouping, and coordinated plan selection. 

Given the number of nodes per layer in DNNs $W$, the state dimension $|s|$, and the action space $|A|$, the computational complexity of DNN policy inference is approximately $O(C_{\text{dnn}}(A)) = O(|A| \cdot W + W^2 + |s| \cdot W)$~\cite{omoniwa2023communication}. Let $C_{\text{grp}}(K,I)$ denotes the cost of dividing $K$ supplied candidate plans into $I$ groups. For brevity, we define the high-level decision costs of HRL and HRCL as $C^\text{HRL}_\text{high} = C_{\text{grp}}(K,I) + C_{\text{dnn}}(I+K/I)$ and $C^\text{HRCL}_\text{high} = C_{\text{grp}}(K,I) + C_{\text{dnn}}(I \cdot M)$ respectively. MAPPO~\cite{nayak2023scalable} directly selects among $K$ candidate plans and therefore has a learning complexity of $O\Big(\mathcal{E} \cdot T \cdot U \cdot C_{\text{dnn}}(K)\Big)$. HRL~\cite{jendoubi2023multi} reduces the individual action spaces through hierarchical selection, resulting in $O\Big(\mathcal{E} \cdot T \cdot U \cdot C^\text{HRL}_\text{high} \Big)$. HRCL instead uses MARL only for high-level guidance and delegates detailed plan selection to decentralized coordination. Its complexity is $O\Big(\mathcal{E} \cdot T \big[ U \cdot C^\text{HRCL}_\text{high} + K/I \cdot L \cdot \log U \big] \Big)$, where the first term represents high-level strategy learning and the second represents low-level tree-based plan selection. In comparison, standalone EPOS performs low-level selection over all $K$ candidate plans with $O\Big(T \cdot K \cdot L \cdot \log U \big] \Big)$. 

The main scalability advantage of HRCL therefore comes from reducing the decision space handled by reinforcement learning from $K$ candidate plans to $I \cdot M$ high-level choices, while restricting low-level coordination to approximately $K/I$ plans per selected group. This advantage becomes more significant as the number of candidate plans increases. Fig.~\ref{fig:complexity} empirically examines the corresponding computational overhead under different numbers of agents and candidate plans. 

\subsection{Effect of different parameters}
The objective of this subsection is to analyze the impact of various parameters within the proposed \emph{HRCL}. The insights derived from this subsection can be used to make informed empirical decisions regarding parameter selection for the evaluation scenarios, including the number of actions $|A|$, agent behavior,$\beta$ and the weights $\sigma$ in reward function. The calculations presented herein can be automated for an instantiated plan-based scenario during hyperparameter optimization. Such optimization is not the focus of this paper. All experiments are performed in the basic synthetic scenario (40 agents, 32 plans per agent and the target signal with $\omega = \pi / 24$).

\cparagraph{Weights in reward function}
Fig.~\ref{fig:weight} shows the effect of weight parameters by varying the $\sigma$ defined in Eq.(\ref{eq:reward}). As $\sigma$ increases from 0 to 1, the weight on discomfort cost in the reward function increases, thereby reducing mean discomfort cost while increasing inefficiency cost in \emph{MAPPO} and \emph{HRCL}. This paper chooses $\sigma = 0.5$ because it gives the best operating point under the selected equal-weight scalarization.

\cparagraph{Agent behavior}
Fig.~\ref{fig:behavior} illustrates the effect of agent behavior by varying the parameter $\beta^u_t$ defined in Eq.(\ref{eq:epos}). As $\beta^u_t$ increases from 0 to 1, agents behave selfishly, and thus reduce the mean discomfort cost while increasing the inefficiency cost. The parameter $\beta^u_t$ controls the agent-level preference used during low-level plan selection. To minimize the selected combined cost, $\beta^u_t = 0.5$ is used in \emph{EPOS} and \emph{HRCL-P}; this selection is specific to the scalarized evaluation.

\cparagraph{Number of actions}
Fig.~\ref{fig:actions} shows the cost comparison of the number of actions of the proposed methods. In \emph{HRCL-P}, when the number of plan groups increases, 
agents can accurately find the plans with lower discomfort cost, as shown in Fig.\ref{fig:frequency_group}, but maintain inefficiency cost. The average frequency denotes the total number of actions taken by all agents over all time periods divided by the number of sub-ranges covered by each group/behavior range. In \emph{HRCL-B}, when the number of behavior ranges increases, agents can sample from a narrow range of behaviors and identify a favorable value, see Fig.\ref{fig:frequency_behavior}. However, with more plan groups and behavior ranges, the costs of these methods increase since high action space makes the training inefficient (see the convergence of \emph{MAPPO} in Fig.~\ref{fig:training}).

\begin{figure}[!t]
    \centering
    \includegraphics[width=\linewidth]{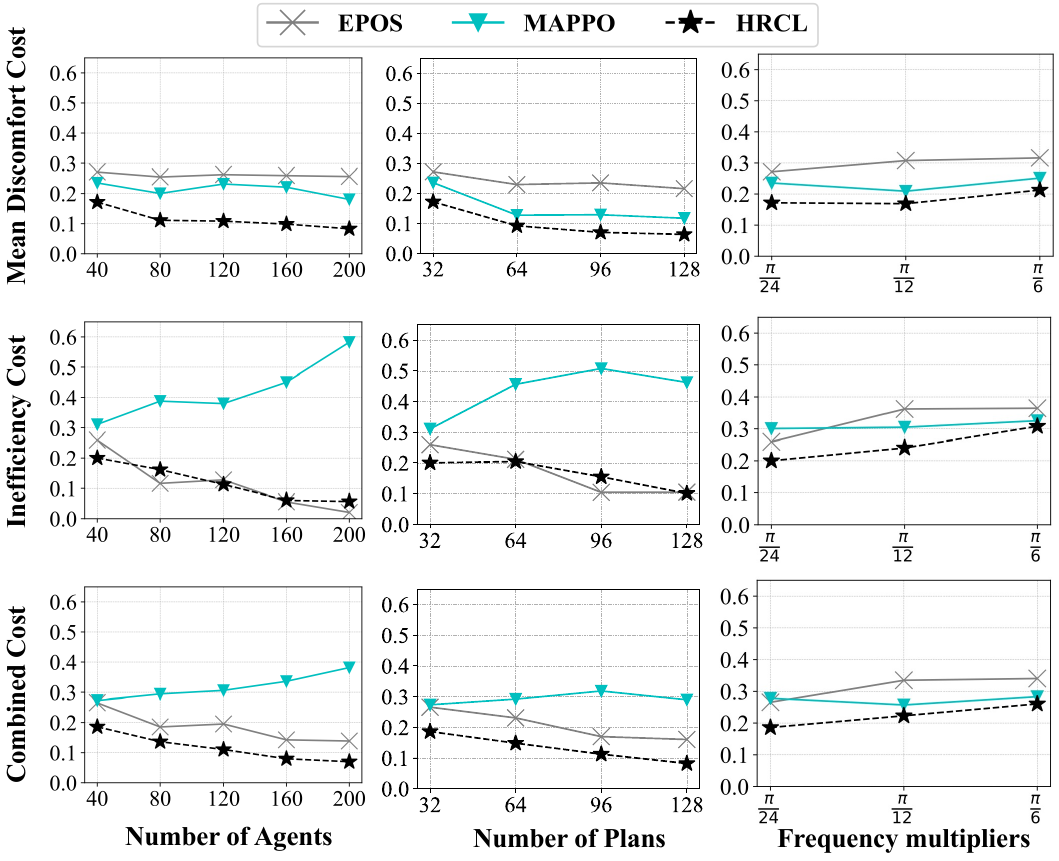}
    \caption{Cost comparison in complex synthetic scenarios: (1) Changing the number of agents from $40$ to $200$ and fixing $32$ plans per agent, $16$ time periods and target complexity; (2) Changing the number of plans per agent from $40$ to $160$ and fixing $40$ agents, $16$ time periods and target complexity; (3) Changing the frequency multiple of cosine waves from $\pi/24$ to $\pi/6$ and fixing $40$ agents, $32$ plans per agent, and $16$ time periods.}
    \label{fig:gaussian_complex}
\end{figure}

\subsection{Evaluation on complex synthetic scenario}
\emph{HRCL} is evaluated using complex synthetic scenario by varying number of agents, number of plans and complexity of target tasks, as shown in Fig.~\ref{fig:gaussian_complex}. 

\cparagraph{Number of agents}
If the number of agents $U$ increases from $40$ to $200$, there are more agents available to effectively coordinate and reach the target tasks. Therefore, the inefficiency cost of \emph{HRCL} drops significantly by around $75\%$. In contrast, \emph{MAPPO} has higher inefficiency cost with more agents due to the expanded state space, which hinders training efficiency. Moreover, \emph{HRCL} strategically coordinates agents to choose plans with low discomfort cost, achieving a $24.34\%$ lower combined cost than \emph{EPOS}. The results illustrate that \textit{high number of agents significantly decreases the inefficiency cost of methods using collective learning}.

\cparagraph{Number of plans}
If the number of plans per agent $K$ increases from $40$ to $160$, each agent has more options to respond to the complex environments. Thus, the combined costs of \emph{HRCL} decreases by $55.91\%$ as the number of plans increases, while still being lower than \emph{EPOS} by $48.75\%$. Unlike \emph{MAPPO}, where agents choose from a significantly larger action space, leading to increased training complexity, the action space of \emph{HRCL} remains constant, contributing to its lower mean discomfort ($45.94\%$) and inefficiency costs ($64.26\%$). The results illustrate that \textit{high number of plans decreases the discomfort and inefficiency cost of methods using collective learning}.

\cparagraph{Complexity of target tasks}
If the frequency multiplier increases from $\pi/24$ to $\pi/6$, agents struggle to achieve the collective goal. This leads to a linear increase of both mean discomfort and inefficiency cost in \emph{EPOS}. In contrast, \emph{MAPPO} keeps the inefficiency cost relatively low and constant since it learns to effectively explore and select plans close to the cosine waveform with high frequency. Via reinforcement learning, the proposed \emph{HRCL} achieves a combined cost that is $33.44\%$ lower than \emph{EPOS}. The results illustrate that \textit{high complexity of target task cannot significantly increase the discomfort and inefficiency costs of MARL methods}.


\subsection{Evaluation on energy self-management} \label{sec:energy_manage}
We study the energy application scenario by using the dataset derived by energy disaggregation of the simulated zonal power transmission in the Pacific Northwest Smart Grid Demonstrations Project~\cite{pournaras2017self}. The plans of each consumer (agent) denotes different energy consumption patterns over 12 hours in a day. The system objective (i.e., \textit{inefficiency cost}) aims to minimize the mismatch (RMSE) between the total energy demand by consumers and a zero target vector in order to mitigate power peaks and reduce the risk of blackouts~\cite{pournaras2018decentralized}. The \textit{discomfort cost} of each plan is the amount of minutes shifted compared to the original demand. More details about the dataset are illustrated in Appendix. 

\begin{figure}[!t]
    \centering
    \includegraphics[width=0.70\linewidth]{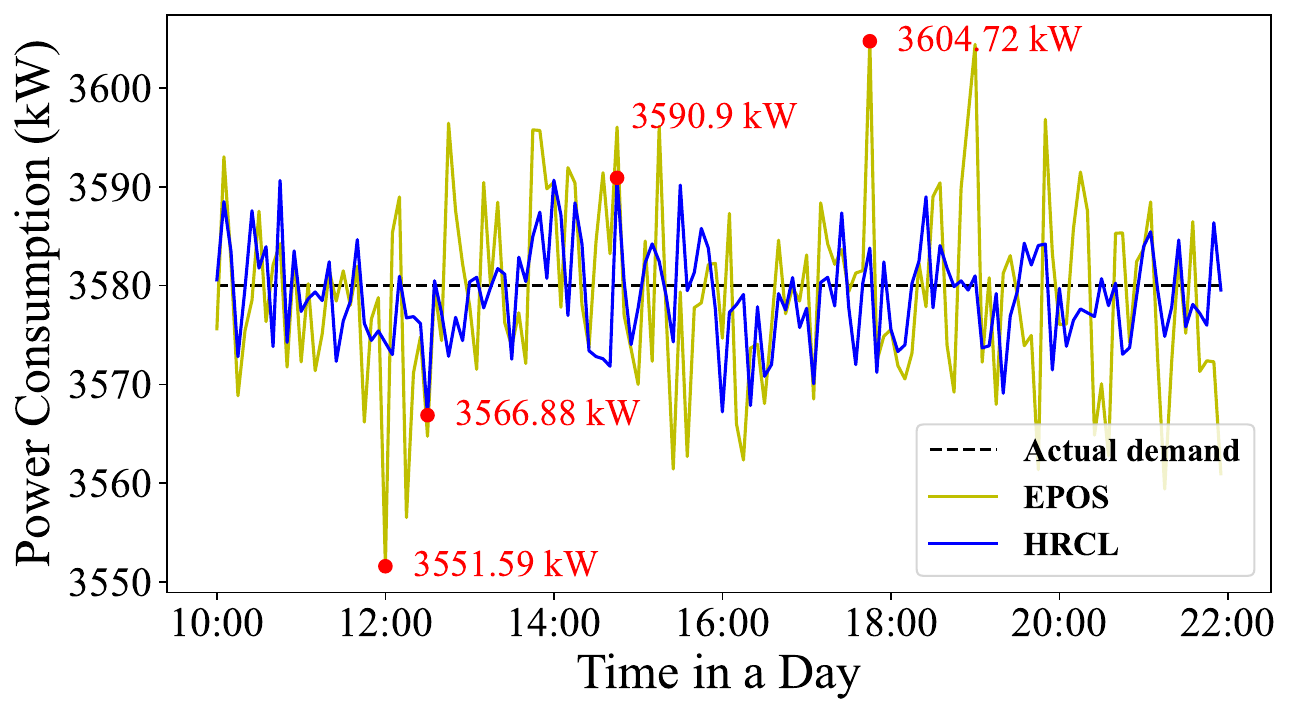}
    \caption{The comparison of total energy demands per time among all methods that show max and min power peak over all periods.}
    \label{fig:energy_response}
\end{figure}

Fig.~\ref{fig:energy_cost} shows that the proposed methods outperform \emph{MAPPO} in terms of inefficiency cost, achieving approximately $36.03\%$ lower combined cost, particularly due to the efficient coordination through multi-agent collective learning. Compared to \emph{EPOS}, \emph{HRCL} reduces max power peaks by $13.82$kW and enhances min power peaks by $15.29$kW, see Fig.~\ref{fig:energy_response}, thereby improving the power grid stability. Among all methods, \emph{HRCL-B} achieves the lowest discomfort and combined costs. Compared to \emph{HRCL-P}, \emph{HRCL-B} helps agents to stay closer to their original energy demands, reducing total deviation time by over $2.8$k minutes across all time periods, significantly enhancing consumer satisfaction. In this scenario, the \emph{grouping behavior ranges} strategy outperforms \emph{grouping plans}. These findings suggest that, in environments with a limited number of plans per agent and where the weighted trade-off between individual and system-wide costs is the primary goal, grouping behavior ranges is more effective than grouping plans.

\begin{figure}[!t]
    \centering
        \includegraphics[width=0.75\linewidth]{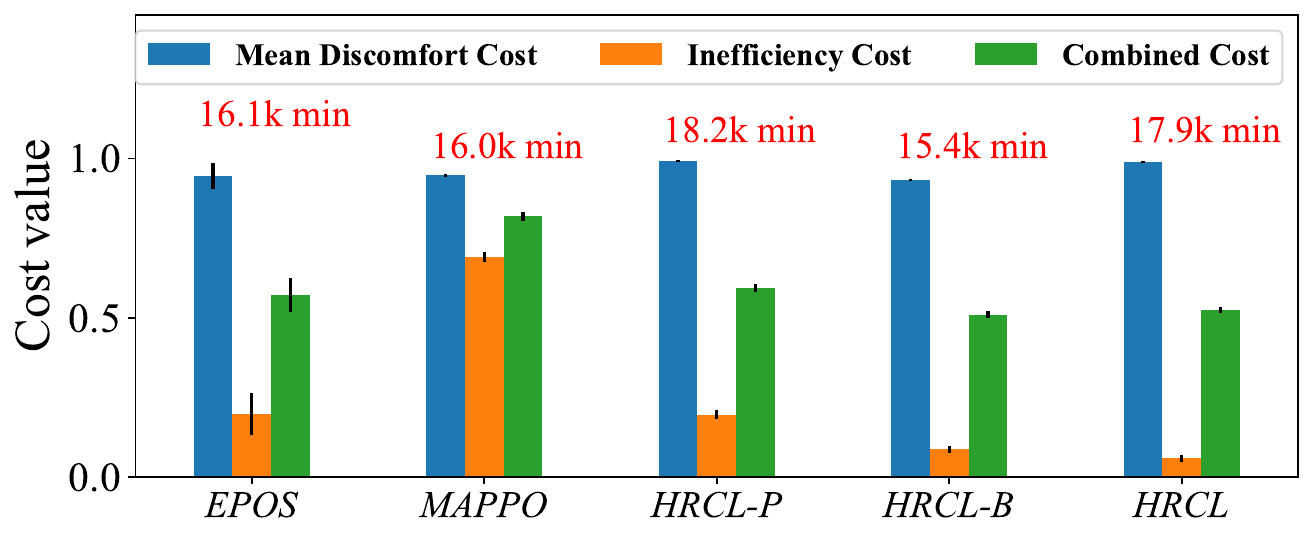}
    \caption{Cost comparison and the total minutes shifted per agent of all methods in the energy self-management scenario ($160$ consumers, $10$ plans per consumer and $16$ time periods, i.e., $16$ days).}
    \label{fig:energy_cost}
\end{figure}

\subsection{Evaluation on drone swarm sensing} \label{sec:drone_sensing}
We study the drone application scenarios by using the dataset derived from navigation and sensing of drone swarms for traffic monitoring in Munich, German~\cite{qin2025strategic}. The sensing tasks prioritize energy efficiency, and thus the \textit{mean discomfort cost} in this scenario represents the energy consumption of drones, whereas \textit{inefficiency cost} represents the sensing inaccuracy that measures the mismatch between overall sensing of drones and sensing requirements (target). The sensing target is estimated from historical vehicle-distribution data using Ordinary Least Squares regression, covering the expected sensing requirements for the current and several future periods. Therefore, the global plan observed by each drone represents the predicted traffic-flow information over these periods, allowing the agent to consider future sensing demands when making current coordination decisions. More details about the dataset are illustrated in Appendix.

This paper employs the \emph{grouping plans} strategy that divides drone plans based on their spatial flight range, i.e., the attribute function is $\phi^u_{kt} = f_\text{s}(p^u_{kt})$. Since each plan represents a navigation and sensing operation of a drone, plans can be grouped by their intended spatial range. Thus, this strategy is significant in a large spatio-temporal sensing missions, where drones can proactively navigate to areas with traffic surges in the future, recharge at nearby base stations, and continue sensing in the next period, even if those areas are currently sparsely populated with vehicles. Such foresight-driven action enhances long-term vehicle detection performance. This method, named as \emph{HRCL-s}, is compared with a baseline method \emph{HRCL-d} that groups the plans based on their discomfort cost.

\begin{figure}[!t]
    \centering
    \includegraphics[width=\linewidth]{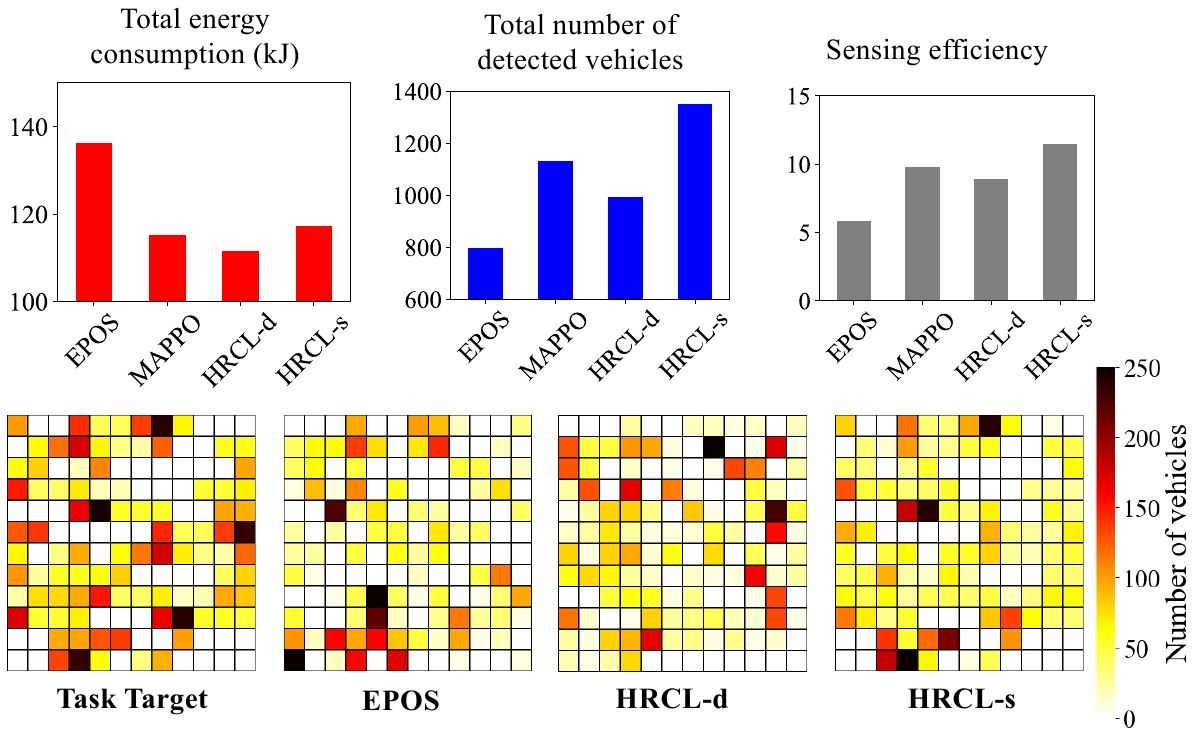}
    \caption{Performance comparison of all methods in the drone swarm sensing scenario ($16$ drones, $64$ plans per drone and $16$ time periods, i.e., $8h$ at total). The metrics include total energy consumption of drones $E$, total number of vehicles detected by drones $V$, and sensing efficiency calculated by $V / E$.}
    \label{fig:drone_basic}
\end{figure}

Fig.~\ref{fig:drone_basic} shows the optimal cost of \emph{HRCL-s} that defines $4$ spatial groups based on directions: north, east, south, and west. This approach achieves relatively lower mean discomfort cost, consuming lower energy ($19.05$ kJ) than \emph{EPOS}. Meanwhile, it efficiently utilizes energy resources for vehicle observation, resulting in the minimum inefficiency cost among all methods. The figure also shows that the number of vehicles detected by drones matches up the real number of vehicles (task target) in the map of $12\times12$ grid cells. Even if the short-term benefits are not evident, \emph{HRCL-s} excels in long-term vehicle detection efficiency and accuracy, outperforming \emph{EPOS} in vehicle detection ($67.47\%$ vs $29.87\%$ of all vehicles in the map). In addition, \emph{HRCL-s} effectively narrows the drone flight ranges to traffic-dense areas and enhances training efficiency, achieving $17.38\%$ high sensing efficiency than \emph{MAPPO}. \emph{HRCL-d} assists drones to choose the route with lower energy consumption, but sacrifice the efficient vehicle detection, resulting in sensing efficiency that is $22.51\%$ lower than \emph{HRCL-s}.

\section{Discussion and New Insights}\label{sec:insights}
The experimental results demonstrate the superior performance of the proposed approach \emph{HRCL}. Several scientific insights on \emph{HRCL} are listed as follows:

\cparagraph{\emph{HRCL} achieves a win-win synthesis of long- and short-term optimization performance improvement in multi-agent systems}
Unlike collective learning approaches such as \emph{EPOS}, \emph{HRCL} leverages DNNs to effectively adapt and tackle complex tasks, leading to $23.69\%$ lower system costs despite higher task complexity. Compared to traditional MARL methods such as \emph{MAPPO}, \emph{HRCL} enables agents to efficiently observe and aggregate the plans of others via a tree communication structure, minimizing the system-wide cost by $31.29\%$. 

\cparagraph{\emph{HRCL} combines the strengths of both high-level strategies to improve scalability and weighted multi-objective trade-offs}
The \emph{grouping plan constraints} strategy in \emph{HRCL} effectively reduces high-level policy complexity by limiting the number of actions that each agent considers, ensuring scalable learning, faster training convergence, and improved cost efficiency even as the numbers of plans and agents increase. Meanwhile, the \emph{grouping behavior ranges} strategy allows each agent to autonomously adjust its agent-level preference used during low-level plan selection, significantly lowering both discomfort and inefficiency costs. This strategy is particularly effective with fewer plans and agents. By integrating these two complementary strategies, \emph{HRCL} balances scalability and solution quality, leading to favorable weighted trade-offs across the evaluated multi-agent scenarios. Moreover, it can use a single strategy in special cases, such as \emph{HRCL-P} for highly complex target tasks and \emph{HRCL-B} for low plan volumes.

\cparagraph{\emph{HRCL} provides a modular coordination layer across plan-based application studies} 
The application-oriented evaluations show that \emph{HRCL} minimizes energy consumption, ensures load balancing, and improves sensing performance in the studied settings. The same coordination architecture can be instantiated across plan-based domains using scenario-specific candidate-plan inputs and attributes. Moreover, \emph{grouping plan constraints} supports flexible grouping criteria, such as plan discomfort costs and semantic information. Thus, \emph{HRCL} is domain-independent at the coordination level while relying on domain-specific candidate-plan construction.

\section{Conclusion and Future Work}\label{sec:conclusion}
This paper studies how reinforcement learning can guide decentralized multi-agent coordination for resource allocation over long time horizons. We propose HRCL, a hierarchical framework that separates long-term strategic learning from short-term decentralized coordination. Instead of using MARL to directly solve the complete combinatorial coordination problem, HRCL learns high-level strategies that guide the existing coordination process. This allows the coordination process to make necessary short-term trade-offs while learning whether these decisions benefit the system over time. The plan grouping strategy reduces the action space by restricting the alternatives considered during coordination, while the behavior strategy guides agents in balancing system-wide efficiency and individual costs. Their combination allows agents to improve coordination decisions over time while maintaining low computational overhead. Experiments on synthetic scenarios demonstrate faster learning and lower system and individual costs than standalone MARL and decentralized coordination approaches. Evaluations on energy self-management and drone swarm sensing further demonstrate the applicability of HRCL and its strategy-specific variants to different resource allocation problems, improving power-load regulation and sensing efficiency. These results show that learning to guide decentralized coordination can turn necessary short-term trade-offs into long-term gains without sacrificing scalability.

Nevertheless, the designed approach can be further improved as follows: (1) Explore fully decentralized training paradigms where agents learn and adapt their policies without relying on centralized coordination~\cite{omidshafiei2017deep}. (2) Study more types of \emph{grouping plans} strategy to adapt to complex real scenarios, e.g., grouping plans that occur during similar time windows (peak or off-peak usage in energy demands).

\bibliographystyle{unsrt}
\bibliography{reference}

@article{teghem2000interactive,
  title={An interactive heuristic method for multi-objective combinatorial optimization},
  author={Teghem, Jacques and Tuyttens, Daniel and Ulungu, Ekunda L},
  journal={Computers \& Operations Research},
  volume={27},
  number={7-8},
  pages={621--634},
  year={2000},
  publisher={Elsevier}
}

@book{deakin2013smart,
  title={Smart cities: governing, modelling and analysing the transition},
  author={Deakin, Mark},
  year={2013},
  publisher={Routledge}
}

@article{liefooghe2014hybrid,
  title={A hybrid metaheuristic for multiobjective unconstrained binary quadratic programming},
  author={Liefooghe, Arnaud and Verel, S{\'e}bastien and Hao, Jin-Kao},
  journal={Applied Soft Computing},
  volume={16},
  pages={10--19},
  year={2014},
  publisher={Elsevier}
}

@article{dong2023decentralised,
  title={Decentralised and cooperative control of multi-robot systems through distributed optimisation},
  author={Dong, Yi and Li, Zhongguo and Zhao, Xingyu and Ding, Zhengtao and Huang, Xiaowei},
  journal={arXiv preprint arXiv:2302.01728},
  year={2023}
}

@inproceedings{picard2023multi,
  title={Multi-agent consensus-based bundle allocation for multi-mode composite tasks},
  author={Picard, Gauthier},
  booktitle={International Conference on Autonomous Agents and Multiagent Systems},
  year={2023}
}

@article{zhang2025neuro,
  title={Neuro-{PLS}: A Generalizable Local Search Framework for Multi-objective Combinatorial Optimization},
  author={Zhang, Haotian and Shi, Jialong and Sun, Jianyong and Zhang, Qingfu and Xu, Zongben},
  journal={IEEE Transactions on Evolutionary Computation},
  year={2025},
  publisher={IEEE}
}

@inproceedings{li2021lifelong,
  title={Lifelong multi-agent path finding in large-scale warehouses},
  author={Li, Jiaoyang and Tinka, Andrew and Kiesel, Scott and Durham, Joseph W and Kumar, TK Satish and Koenig, Sven},
  booktitle={Proceedings of the AAAI Conference on Artificial Intelligence},
  volume={35},
  number={13},
  pages={11272--11281},
  year={2021}
}

@inproceedings{hinrichs2013cohda,
  title={{COHDA}: A combinatorial optimization heuristic for distributed agents},
  author={Hinrichs, Christian and Lehnhoff, Sebastian and Sonnenschein, Michael},
  booktitle={International Conference on Agents and Artificial Intelligence},
  pages={23--39},
  year={2013},
  organization={Springer}
}

@article{chen2018bd,
  title={BD-ADOPT: a hybrid DCOP algorithm with best-first and depth-first search strategies},
  author={Chen, Ziyu and He, Chen and He, Zhen and Chen, Minyou},
  journal={Artificial Intelligence Review},
  volume={50},
  number={2},
  pages={161--199},
  year={2018},
  publisher={Springer}
}

@article{li2020deep,
  title={Deep reinforcement learning for multiobjective optimization},
  author={Li, Kaiwen and Zhang, Tao and Wang, Rui},
  journal={IEEE transactions on cybernetics},
  volume={51},
  number={6},
  pages={3103--3114},
  year={2020},
  publisher={IEEE}
}

@inproceedings{wu2024collaborative,
  title={Collaborative deep reinforcement learning for solving multi-objective vehicle routing problems},
  author={Wu, Yaoxin and Fan, Mingfeng and Cao, Zhiguang and Gao, Ruobin and Hou, Yaqing and Sartoretti, Guillaume},
  booktitle={23rd International Conference on Autonomous Agents and Multiagent Systems, AAMAS 2024},
  pages={1956--1965},
  year={2024},
  organization={International Foundation for Autonomous Agents and Multiagent Systems (IFAAMAS)}
}

@inproceedings{barrett2020exploratory,
  title={Exploratory combinatorial optimization with reinforcement learning},
  author={Barrett, Thomas and Clements, William and Foerster, Jakob and Lvovsky, Alex},
  booktitle={Proceedings of the AAAI conference on artificial intelligence},
  volume={34},
  number={04},
  pages={3243--3250},
  year={2020}
}

@inproceedings{cappart2021combining,
  title={Combining reinforcement learning and constraint programming for combinatorial optimization},
  author={Cappart, Quentin and Moisan, Thierry and Rousseau, Louis-Martin and Pr{\'e}mont-Schwarz, Isabeau and Cire, Andre A},
  booktitle={Proceedings of the AAAI Conference on Artificial Intelligence},
  volume={35},
  number={5},
  pages={3677--3687},
  year={2021}
}

@article{chen2022ptde,
  title={PTDE: Personalized training with distilled execution for multi-agent reinforcement learning},
  author={Chen, Yiqun and Mao, Hangyu and Mao, Jiaxin and Wu, Shiguang and Zhang, Tianle and Zhang, Bin and Yang, Wei and Chang, Hongxing},
  journal={arXiv preprint arXiv:2210.08872},
  year={2022}
}

@article{andrychowicz2016learning,
  title={Learning to learn by gradient descent by gradient descent},
  author={Andrychowicz, Marcin and Denil, Misha and Gomez, Sergio and Hoffman, Matthew W and Pfau, David and Schaul, Tom and Shillingford, Brendan and De Freitas, Nando},
  journal={Advances in neural information processing systems},
  volume={29},
  year={2016}
}

@article{yu2022surprising,
  title={The surprising effectiveness of {PPO} in cooperative multi-agent games},
  author={Yu, Chao and Velu, Akash and Vinitsky, Eugene and Gao, Jiaxuan and Wang, Yu and Bayen, Alexandre and Wu, Yi},
  journal={Advances in neural information processing systems},
  volume={35},
  pages={24611--24624},
  year={2022}
}

@article{jendoubi2023multi,
  title={Multi-agent hierarchical reinforcement learning for energy management},
  author={Jendoubi, Imen and Bouffard, Fran{\c{c}}ois},
  journal={Applied Energy},
  volume={332},
  pages={120500},
  year={2023},
  publisher={Elsevier}
}

@article{wong2023deep,
  title={Deep multiagent reinforcement learning: Challenges and directions},
  author={Wong, Annie and B{\"a}ck, Thomas and Kononova, Anna V and Plaat, Aske},
  journal={Artificial Intelligence Review},
  volume={56},
  number={6},
  pages={5023--5056},
  year={2023},
  publisher={Springer}
}

@inproceedings{han2023model,
  title={Model-based sparse communication in multi-agent reinforcement learning},
  author={Han, Shuai and Dastani, Mehdi and Wang, Shihan},
  booktitle={Seventeenth European Workshop on Reinforcement Learning},
  year={2023}
}

@inproceedings{li2024q,
  title={A Q-learning evolutionary multiobjective framework for multiobjective optimization with separable and interacting variables},
  author={Li, Hui and Tang, Yanhui and Shui, Yuxiang and Sun, Jianyong},
  booktitle={2024 IEEE Congress on Evolutionary Computation (CEC)},
  pages={1--8},
  year={2024},
  organization={IEEE}
}

@inproceedings{tilak2010decentralized,
  title={Decentralized and partially decentralized reinforcement learning for distributed combinatorial optimization problems},
  author={Tilak, Omkar and Mukhopadhyay, Snehasis},
  booktitle={2010 Ninth International Conference on Machine Learning and Applications},
  pages={389--394},
  year={2010},
  organization={IEEE}
}

@inproceedings{majeed2021exact,
  title={Exact reduction of huge action spaces in general reinforcement learning},
  author={Majeed, Sultan J and Hutter, Marcus},
  booktitle={Proceedings of the AAAI Conference on Artificial Intelligence},
  volume={35},
  number={10},
  pages={8874--8883},
  year={2021}
}

@article{machado2023temporal,
  title={Temporal abstraction in reinforcement learning with the successor representation},
  author={Machado, Marlos C and Barreto, Andre and Precup, Doina and Bowling, Michael},
  journal={Journal of Machine Learning Research},
  volume={24},
  number={80},
  pages={1--69},
  year={2023}
}

@article{pateria2021hierarchical,
  title={Hierarchical reinforcement learning: A comprehensive survey},
  author={Pateria, Shubham and Subagdja, Budhitama and Tan, Ah-hwee and Quek, Chai},
  journal={ACM Computing Surveys (CSUR)},
  volume={54},
  number={5},
  pages={1--35},
  year={2021},
  publisher={ACM New York, NY, USA}
}

@article{ding2021crowdsourcing,
  title={When crowdsourcing meets unmanned vehicles: Toward cost-effective collaborative urban sensing via deep reinforcement learning},
  author={Ding, Lige and Zhao, Dong and Cao, Mingzhe and Ma, Huadong},
  journal={IEEE Internet of Things Journal},
  volume={8},
  number={15},
  pages={12150--12162},
  year={2021},
  publisher={IEEE}
}

@article{ma2024efficient,
  title={Efficient and scalable reinforcement learning for large-scale network control},
  author={Ma, Chengdong and Li, Aming and Du, Yali and Dong, Hao and Yang, Yaodong},
  journal={Nature Machine Intelligence},
  volume={6},
  number={9},
  pages={1006--1020},
  year={2024},
  publisher={Nature Publishing Group UK London}
}

@article{bernstein2002complexity,
  title={The complexity of decentralized control of Markov decision processes},
  author={Bernstein, Daniel S and Givan, Robert and Immerman, Neil and Zilberstein, Shlomo},
  journal={Mathematics of operations research},
  volume={27},
  number={4},
  pages={819--840},
  year={2002},
  publisher={INFORMS}
}

@article{omoniwa2023communication,
  title={Communication-enabled deep reinforcement learning to optimise energy-efficiency in {UAV}-assisted networks},
  author={Omoniwa, Babatunji and Galkin, Boris and Dusparic, Ivana},
  journal={Vehicular Communications},
  volume={43},
  pages={100640},
  year={2023},
  publisher={Elsevier}
}

@article{lowe2017multi,
  title={Multi-agent actor-critic for mixed cooperative-competitive environments},
  author={Lowe, Ryan and Wu, Yi I and Tamar, Aviv and Harb, Jean and Pieter Abbeel, OpenAI and Mordatch, Igor},
  journal={Advances in neural information processing systems},
  volume={30},
  year={2017}
}

@inproceedings{nayak2023scalable,
  title={Scalable multi-agent reinforcement learning through intelligent information aggregation},
  author={Nayak, Siddharth and Choi, Kenneth and Ding, Wenqi and Dolan, Sydney and Gopalakrishnan, Karthik and Balakrishnan, Hamsa},
  booktitle={International conference on machine learning},
  pages={25817--25833},
  year={2023},
  organization={PMLR}
}

@inproceedings{xu2023haven,
  title={Haven: Hierarchical cooperative multi-agent reinforcement learning with dual coordination mechanism},
  author={Xu, Zhiwei and Bai, Yunpeng and Zhang, Bin and Li, Dapeng and Fan, Guoliang},
  booktitle={Proceedings of the AAAI Conference on Artificial Intelligence},
  volume={37},
  number={10},
  pages={11735--11743},
  year={2023}
}

@inproceedings{omidshafiei2017deep,
  title={Deep decentralized multi-task multi-agent reinforcement learning under partial observability},
  author={Omidshafiei, Shayegan and Pazis, Jason and Amato, Christopher and How, Jonathan P and Vian, John},
  booktitle={International Conference on Machine Learning},
  pages={2681--2690},
  year={2017},
  organization={PMLR}
}

@article{iqbal2022alma,
  title={{ALMA}: Hierarchical learning for composite multi-agent tasks},
  author={Iqbal, Shariq and Costales, Robby and Sha, Fei},
  journal={Advances in neural information processing systems},
  volume={35},
  pages={7155--7166},
  year={2022}
}

@article{zang2023automatic,
  title={Automatic grouping for efficient cooperative multi-agent reinforcement learning},
  author={Zang, Yifan and He, Jinmin and Li, Kai and Fu, Haobo and Fu, Qiang and Xing, Junliang and Cheng, Jian},
  journal={Advances in neural information processing systems},
  volume={36},
  pages={46105--46121},
  year={2023}
}

@inproceedings{liu2026hcpo,
  title={HCPO: Hierarchical Conductor-Based Policy Optimization in Multi-Agent Reinforcement Learning},
  author={Liu, Zejiao and Tu, Junqi and Hong, Yitian and Xiong, Luolin and Jin, Yaochu and Tang, Yang and Li, Fangfei},
  booktitle={Proceedings of the AAAI Conference on Artificial Intelligence},
  volume={40},
  number={35},
  pages={29564--29572},
  year={2026}
}

@article{pournaras2018decentralized,
  title={Decentralized collective learning for self-managed sharing economies},
  author={Pournaras, Evangelos and Pilgerstorfer, Peter and Asikis, Thomas},
  journal={ACM Transactions on Autonomous and Adaptive Systems (TAAS)},
  volume={13},
  number={2},
  pages={1--33},
  year={2018},
  publisher={ACM New York, NY, USA}
}

@article{pournaras2017self,
  title={Self-regulating supply--demand systems},
  author={Pournaras, Evangelos and Yao, Mark and Helbing, Dirk},
  journal={Future Generation Computer Systems},
  volume={76},
  pages={73--91},
  year={2017},
  publisher={Elsevier}
}

@article{Pournaras2023,
author = "Evangelos Pournaras",
title = "{Agent-based Planning Portfolio}",
year = "2019",
month = "4",
url = "https://figshare.com/articles/dataset/Agent-based_Planning_Portfolio/7806548",
doi = "10.6084/m9.figshare.7806548.v6",
note = {DOI: https://doi.org/10.6084/m9.figshare.7806548.v6}
}

@inproceedings{majumdar2023discrete,
  title={Discrete-choice multi-agent optimization: Decentralized hard constraint satisfaction for smart cities},
  author={Majumdar, Srijoni and Qin, Chuhao and Pournaras, Evangelos},
  booktitle={International conference on autonomous agents and multiagent systems},
  pages={60--76},
  year={2023},
  organization={Springer}
}

@article{qin2025strategic,
  title={Strategic coordination of drones via short-term distributed optimization and long-term reinforcement learning},
  author={Qin, Chuhao and Pournaras, Evangelos},
  journal={Applied Soft Computing},
  pages={114270},
  year={2025},
  publisher={Elsevier}
}

@misc{cpsatlp,
  title = {{CP-SAT}},
  author = {Laurent Perron and Frédéric Didier},
  organization = {Google},
  url = {https://developers.google.com/optimization/cp/cp_solver/},
  note = {Google OR-Tools, version 9.12},
  year = {2025}
}


\vfill

\end{document}